\documentclass[twocolumn,aps,pre,floatfix,superscriptaddress]{revtex4-2}
\usepackage{psfrag,epsfig,amsfonts,amssymb,amsmath,wasysym,bm}
\usepackage{bbold}
\usepackage{color}
\usepackage{titlesec}

\usepackage[inline]{enumitem} 
\usepackage{hyperref}

\newcommand{\<}{\langle}	
\renewcommand{\>}{\rangle}	
\newcommand{\ket}[1]{\lvert #1 \rangle} 	
\newcommand{\bra}[1]{\langle #1 \rvert}	
\newcommand{\ketN}[1]{\lvert #1 \rangle} 	
\newcommand{\braN}[1]{\langle #1 \rvert}	
\newcommand{\mc}{\mathcal}	
\newcommand{\sUp}{\uparrow}		
\newcommand{\sDn}{\downarrow}	
 
\renewcommand{\Re}{\operatorname{Re}}
\renewcommand{\Im}{\operatorname{Im}}

\newcommand{\e}{\mathrm{e}}		
\newcommand{\I}{\mathrm{i}}		

\newcommand{\id}{\mathbb 1}
\newcommand{\RR}{{\mathbb R}}
\newcommand{\CC}{{\mathbb C}}

\newcommand{\ZZ}{{\mathbb Z}}

\newcommand{\tr}{\operatorname{tr}}

\newcommand{\mic}{\mathrm{mc}}

\newcommand{\dof}{N_{\mathrm{dof}}}

\newcommand{\Autav}{\bar A_0}
\newcommand{\Auth}{A_{\mathrm{th}}}

\newcommand{\varV}{\tilde v}
\newcommand{\varVFT}{v}

\providecommand{\av}[1]{\mathbb{E}[#1]}
\providecommand{\avv}[1]{\mathbb{E}\!\left[#1\right]}

\newcommand{\lav}[2]{[#1]_{#2}}

\renewcommand{\d}{\mathrm{d}} 
\newcommand{\T}{\mathsf{T}}     
\newcommand{\lmat}{\left( \begin{matrix}}	
\newcommand{\rmat}{\end{matrix} \right)}	
\newcommand{\str}{\operatorname{str}} 
\newcommand{\dos}{D_0}

\newcommand{\tResp}{t_{\mathrm{resp}}}
\newcommand{\tStall}{t_{\mathrm{stall}}}
\newcommand{\tHeat}{t_{\mathrm{heat}}}

\newcommand{\Msz}{s^z_{\mathrm{stag}}}

\newcommand{\HFloq}{H_{\mathrm F}}

\newcommand{\mref}[1]{\ref{#1}}
\newcommand{\meqref}[1]{\eqref{#1}}

\definecolor{darkgreen}{RGB}{32, 150, 32}

\definecolor{orange}{RGB}{210,170,32}

\begin{document}

\title{Stalled response near thermal equilibrium in periodically driven 
systems}

\author{Lennart Dabelow}
\affiliation{RIKEN Center for Emergent Matter Science (CEMS), Wako, Saitama 351-0198, Japan}
\affiliation{School of Mathematical Sciences, Queen Mary University of London, London E1 4NS, UK}

\author{Peter Reimann}
\affiliation{Faculty of Physics, 
Bielefeld University, 
33615 Bielefeld, Germany}
\date{\today}

\begin{abstract}
The question of how
systems respond to perturbations
is ubiquitous in physics.
Predicting this response for large classes of systems
becomes particularly challenging
if
many degrees of
freedom are involved and
linear response theory cannot be applied.
Here, we consider 
isolated 
many-body quantum systems which 
either start out far from equilibrium and then thermalize,
or find themselves near thermal equilibrium from the outset.
We show that time-periodic perturbations 
of moderate strength, in the sense that
they do not heat up the system too 
quickly, give rise to the following
phenomenon of stalled response:
While the driving 
usually causes quite considerable reactions 
as long as the unperturbed system is far from 
equilibrium, the driving
effects 
are strongly suppressed
when the unperturbed system approaches thermal 
equilibrium.
Likewise, for systems
prepared near thermal equilibrium,
the response to the driving is 
barely noticeable
right from the beginning.
Numerical results are complemented by
a quantitatively accurate analytical
description
and by simple qualitative arguments.
\end{abstract}

\maketitle


Understanding the effect of time-dependent perturbations on many-body quantum 
systems is a fundamental problem of immediate practical relevance.
Examples include the implementation of cold-atom \cite{
blo08, blo12, bla12, lan15,
ued20,
geo14}
and polarization-echo 
\cite{geo14,
pen21,
bea21}
experiments,
or the control of general-purpose quantum computers and simulators \cite{nie10, blo12, bla12, geo14}.
Periodic driving, in particular, has been exploited to design 
so-called time crystals
\cite{han22}
and various meta materials with unforeseen topological 
and dynamical properties, whose exploration has only just begun \cite{
hol16,
oka19, moe17,
wei21}.

In this context, the majority of previous studies focused on the long-time 
behavior and, in particular, on the properties of the so-called Floquet Hamiltonian.
A key aspect of such an approach is that it can only capture the 
actual behavior of the periodically driven system stroboscopically in time, 
i.e., at integer multiples of the driving period, whereas
the possibly still very rich behavior 
in between those discrete time points
remains inaccessible.
For instance,
the stroboscopic dynamics
may appear
nearly
stationary
even though the full, continuous dynamics
still
exhibits oscillations with
large amplitudes.

We
adopt a complementary perspective and explore
the continuously time-resolved response
on short-to-intermediate time scales.
Intuitively, one might naturally
expect that
periodic forcing
leads to a clearly noticeable change of the observable 
properties if its strength and period are of the same order
as the 
main
intrinsic energy and time scales of the 
undriven
system.

In this work, we
show that such a fairly pronounced response is indeed observed for isolated
many-body systems that are far away from thermal equilibrium.
Our main discovery, however, is that 
this intuitively expected
response is strongly suppressed near thermal 
equilibrium,
at least as long as heating effects of the driving remain negligible.
We dub this 
phenomenon 
``stalled response'' 
in view of its two principal manifestations:
For a system that is prepared far away from equilibrium,
the observable response dies out as 
soon as the 
corresponding undriven 
reference system 
approaches thermal equilibrium.
Similarly, when the system already starts out in thermal equilibrium,
the driving is barely noticeable 
right from the beginning.
In both cases, it is only at much later times that the driving effects may reappear in the form of very slow heating.
Besides numerical evidence from several examples,
we
support our general prediction of stalled response
near thermal equilibrium with
simple heuristic arguments
and
with an analytical theory for
large classes of many-body systems.
Remarkably,
we can also identify
the main 
qualitative 
signatures of such a stalled response
behavior 
in data from a very recent NMR experiment \cite{bea21}.

\begin{figure*}
\includegraphics[scale=1]{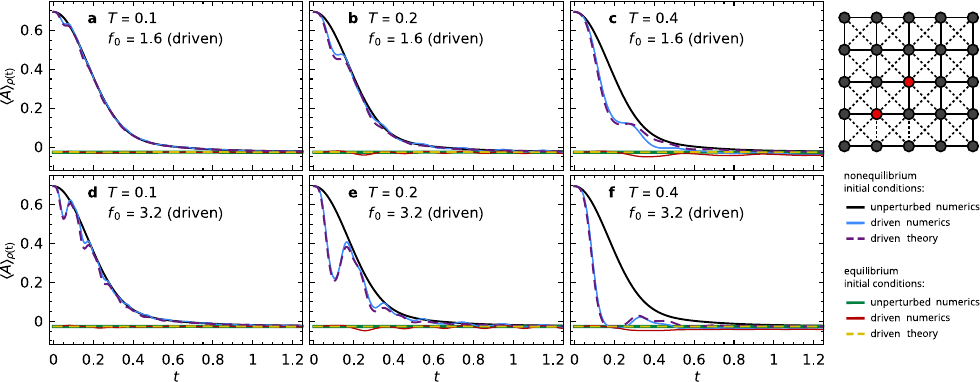}
\caption{Stalled response in a $5 \times 5$ lattice spin system.
Time-dependent expectation values $\< A \>_{\!\rho(t)}$ of the magnetization 
correlation $A = \sigma^z_{2,2} \, \sigma^z_{3,3}$ 
are shown for a periodically driven
system (see sketch)
with 
Hamiltonian \eqref{eq:H}, \eqref{eq:Spin5x5:H0}, \eqref{eq:Spin5x5:V},
\eqref{eq:DrivingProtocol:Sin}.
Solid black and blue lines: numerical results for non-equilibrium
initial conditions~(\ref{eq:InitState}) with
$Q = \pi^+_{2,2} \pi^+_{3,3}$, for driving amplitudes 
$f_0=0$ (unperturbed, black) and
for driving periods $T$ and amplitudes
$f_0$ as indicated in each panel (driven, blue).
Solid green and red lines: same but for equilibrium
initial conditions~(\ref{eq:InitState}) with $Q = \id$.
Dashed lines: corresponding
theoretical predictions~\eqref{eq:TypTimeEvo}, 
adopting the numerically obtained unperturbed behavior $\< A \>_{\!\rho_0(t)}$,
squared response function $|\gamma_t(t)|^2$ 
(by numerical integration of~\eqref{eq:ResProfEq}), 
and thermal equilibrium value $\Auth = -0.026$ (see below Eq.~(\ref{eq:DrivingProtocol:Sin})).}
\label{fig:Spin5x5}
\end{figure*}

\section*{RESULTS}

We
consider periodically driven many-body
systems with Hamiltonians
\begin{equation}
\label{eq:H}
	H(t) = H_0 + f(t) \, V \,,
\end{equation}
where $H_0$ models some unperturbed reference system,
$V$ is a perturbation operator,
and $f(t) = f(t + T)$ is a (scalar) function with period $T$.

As usual,
the expectation value of an observable (Hermitian operator)
$A$ then follows as
\begin{equation}
\label{eq:TimeEvo}
	\< A \>_{\!\rho(t)} := 
\tr\{ \rho(t) A \}
\,,
\end{equation}
where $\rho(t) := \mc U(t) \rho(0) \mc U^\dagger(t)$ is the
(pure or mixed) system state at time $t$ 
if the initial condition was $\rho(0)$,
and the propagator $\mc U(t)$ satisfies $\frac{\d}{\d t} \mc U(t) = -\I H(t) \mc U(t)$ 
and $\mc U(0) = \id$ (identity operator).
Likewise, the unperturbed system 
starts out from the same initial state $\rho(0)$,
and then evolves into $\rho_0(t)$ under the 
time-independent Hamiltonian $H_0$, yielding
expectation values 
$\< A \>_{\!\rho_0(t) }:= \tr\{ \rho_0(t) A \}$.
Accordingly, the system's response to the driving
is monitored by the deviations of
$\< A \>_{\!\rho(t)}$ from $\< A \>_{\!\rho_0(t) }$.

\subsection*{Phenomenology}

To illustrate the announced phenomenon of stalled response,
we first present a
numerical example
in Fig.~\ref{fig:Spin5x5}.
Its specific choice
is
mainly motivated by the fact that it
will 
admit a direct comparison with 
our analytical theory (presented below)
without any free fit parameter.
Further examples will be 
provided 
later.

As sketched in Fig.~\ref{fig:Spin5x5}, 
we consider an $L \times L$ spin-$\frac{1}{2}$ lattice
with $L=5$ and open boundary conditions,
where nearest neighbors are coupled by Heisenberg 
terms in the unperturbed system
(solid links in the sketch),
\begin{equation}
\label{eq:Spin5x5:H0}
	H_0 := \sum_{i,j = 1}^{L-1} \bm\sigma_{i,j} \cdot (\bm\sigma_{i+1,j} + \bm\sigma_{i,j+1}) \,.
\end{equation}
The vector $\bm\sigma_{i,j} = (\sigma^x_{i,j}, \sigma^y_{i,j}, \sigma^z_{i,j})$ 
collects the Pauli matrices acting on site $(i,j)$.
The 
perturbation
additionally introduces spin-flip terms in the $z$ direction between 
next-nearest neighbors
(dashed links in the sketch),
\begin{equation}
\label{eq:Spin5x5:V}
	V := \sum_{i,j=1}^{L-1} \sum_{\alpha = x,y} (\sigma^\alpha_{i,j} \sigma^\alpha_{i+1,j+1} 
	+ \sigma^\alpha_{i+1,j} \sigma^\alpha_{i,j+1}) \,.
\end{equation}
Since the 
magnetization 
$S^z := \sum_{i,j} \sigma^z_{i,j}$
commutes with both $H_0$ and $V$,
we focus on
one of the two largest subsectors, namely
the one with eigenvalue $-1$
for $S^z$.

To prepare the system out of equilibrium,
we fix the spins at sites $(2,2)$ and $(3,3)$ 
in the ``up'' state 
(red in the 
sketch in Fig.~\ref{fig:Spin5x5})
and orient all other 
spins randomly.
To
obtain a well-defined energy,
we additionally emulate a macroscopic energy measurement by acting with a 
Gaussian filter \cite{pre95, gar13, ste14} of a target mean energy $E = -12$ 
and standard deviation $\Delta E = 4$ on the so-defined state.
Formally, the initial condition can thus be expressed as
$\rho(0) = \ket\psi \! \bra\psi$ with
\begin{equation}
\label{eq:InitState}
	\ket\psi \propto \e^{-(H_0 - E)^2 / 4 \Delta E^2} \, 
	Q \,
	\ket\phi \,,
\end{equation}
where $\ket\phi$ is a Haar-random state in the $S^z = -1$ sector.
The projector $Q := \pi^+_{2,2} \, \pi^+_{3,3}$ with $\pi^\pm_{i,j} := (1 \pm \sigma^z_{i,j})/2$ 
enforces $\sigma^z_{2,2} = \sigma^z_{3,3} = 1$,
and this deflection is only weakly reduced by the subsequent 
Gaussian energy filter
(cf.\ Fig.~\ref{fig:Spin5x5}).
From a different viewpoint, the situation may also be seen as a 
small
non-equilibrium system in contact with a 
large
thermal bath (red and black vertices, respectively, in the sketch).

Accordingly, an obvious choice for the considered observable is the 
correlation between the initially disequilibrated sites, 
$A = \sigma^z_{2,2} \, \sigma^z_{3,3}$.

Incidentally,
the ground-state energy of $H_0$ from~\eqref{eq:Spin5x5:H0} is approximately $-60$,
whereas the infinite-temperature state has an energy of approximately $-1$.
Hence, 
our 
choice of the
target energy $E = -12$ 
should be 
reasonably 
generic
and corresponds, as detailed in 
Supplementary Note~2.2,
to an inverse temperature $\beta \approx 0.08$.
Further examples for different target energies/temperatures can also
be found in
Supplementary Note~2.2.

In Fig.~\ref{fig:Spin5x5} we present 
numerical results,
obtained by Suzuki-Trotter propagation,
for the unperturbed system $H_0$
and for a sinusoidally driven 
system (\ref{eq:H}) with
\begin{equation}
\label{eq:DrivingProtocol:Sin}
	f(t) = f_0 \, \sin(2\pi t / T)
	\,,
\end{equation}
yielding the solid black and blue lines,
respectively.
 
The key observation is that the driven (blue)
and undriven (black)
expectation values in Fig.~\ref{fig:Spin5x5} 
differ quite notably during the
initial relaxation of the unperturbed system, but they
become (nearly) indistinguishable 
upon approaching their (almost) steady long-time values.
Moreover, both long-time values
agree very well with the thermal expectation value $\Auth \simeq -0.026$,
obtained numerically by evaluating 
$A = \sigma^z_{2,2} \, \sigma^z_{3,3}$
in 
the 
microcanonical ensemble of the 
unperturbed
system.
In other words, 
the perturbations by the periodic driving get stalled upon 
thermalization of the undriven system.

To further
highlight
this phenomenon,
let us also consider the analogous equilibrium
initial conditions 
with $Q = \id$ in~\eqref{eq:InitState}.
Hence,
the initial state
populates the same energy window as
in the 
nonequilibrium setting,
but the observable expectation values now (approximately) 
assume the pertinent thermal equilibrium values
\cite{pre95, gar13, ste14}.
The solid green and red lines in Fig.~\ref{fig:Spin5x5} 
illustrate the so-obtained numerical results
for the unperturbed and the
driven
system.
In particular, the initial expectation value
is now
very close to the thermal equilibrium 
value $\Auth \simeq -0.026$.
Moreover, the effects of the driving are indeed barely 
noticeable, 
and are 
even expected to 
become still smaller for larger system sizes, 
as detailed in
Supplementary Note~2.3.

The 
bottom line of all these numerical findings is that
the same system exhibits a quite significant 
response to the periodic driving away from 
thermal equilibrium, but
hardly shows any 
reaction 
to the same driving 
as the unperturbed system approaches thermal equilibrium,
or if it already started out near thermal equilibrium
(stalled response). 

Note that the driving amplitudes in Fig.~\ref{fig:Spin5x5}
are far outside the linear response regime,
as can be inferred, e.g.,
by comparing the blue curves of Figs.~\ref{fig:Spin5x5}c 
and f (see also 
Supplementary Note~2.1).
We also remark that for noncommuting perturbations 
and observables (as in Fig.~\ref{fig:Spin5x5}),
linear response theory generically excludes that there is no response 
at all.
The main challenge is to understand why the non-linear response 
remains so weak at thermal equilibrium.

Likewise, the observable response becomes uninterestingly 
weak for extremely small or large driving periods $T$,
regardless of the initial conditions and their proximity to 
thermal
equilibrium.
Hence, our focus here is on the natural regime of moderate 
$T$ values that are similar to,
or slightly below the relaxation time of the unperturbed system,
where the stalling effect is most pronounced and interesting.
The interplay of the various time scales is 
further elaborated in
Supplementary Note~1.1.

Finally, it is well-established that, 
for sufficiently large times, the driving will ultimately
heat up the system towards 
a thermal steady state with infinite temperature 
\cite{dal14, laz14equilibrium,
mal19heating, pon15manybody, ish18}.
However, it is equally well-established that this
heating may often happen only very slowly,
particularly for sufficiently small driving periods $T$ \cite{mor16, aba17effective, aba15, aba17rigorous}.
Our present stalled response effect
thus complements and substantially extends 
those previous predictions from 
Refs.~\cite{dal14, laz14equilibrium,
mal19heating, pon15manybody, ish18}.

\subsection*{Theory}

Our next goal is to establish an analytical theory 
for reasonably general classes of many-body quantum systems
which explains these numerical findings. 
We start by collecting the basic 
ingredients and assumptions,
then present the main result, 
and finally sketch the derivation.

First,
we focus on initial states $\rho(0)$
with a well-defined macroscopic energy.
Denoting by $E_\mu$ and $\ketN{\mu}$
the eigenvalues and -vectors of the
unperturbed Hamiltonian $H_0$,
this means that non-negligible level populations
$\braN{\mu}\rho(0) \ketN{\mu}$
only occur for energies $E_\mu$ within a sufficiently 
small energy interval $\Delta$,
such that the
density of states
can be 
approximated by a constant
$\dos$
throughout 
$\Delta$.

Second, within this energy interval $\Delta$,
the matrix elements $V_{\mu\nu} := \braN{\mu} V \ketN{\nu}$ 
of the 
perturbation operator $V$
are assumed to exhibit a well-defined 
perturbation profile
\begin{equation}
\label{eq:VarV}
	\varV(E) := \lav{ \lvert V_{\mu\nu} \rvert^2 }{E} \,,
\end{equation}
where $\lav{ \,\cdots }{E}$ denotes a local average over 
matrix elements with $\lvert E_\mu - E_\nu \rvert \approx E$.
The perturbation profile's Fourier transform is denoted as
\begin{equation}
\label{eq:VarVFT}
	\varVFT(t) := \int
	\d E \, \dos
	 \, \varV(E) \, \e^{\I E t} \,.
\end{equation}
In passing, we note that at sufficiently high temperatures,
$v(t)$ can be approximated by the two-point correlation function 
$\< V(t) V \>_{\!\rho_{\mic}} / 2$, where $V(t) := \e^{\I H_0 t} V \e^{-\I H_0 t}$ 
and $\rho_{\mic}$ is the microcanonical ensemble corresponding to the energy interval $\Delta$;
see
Supplementary Note~3
for details.

Third, the time-dependent perturbations $f(t) V$ in (\ref{eq:H})
should not become overly strong compared to $H_0$, 
so that establishing a connection between the unperturbed and 
driven
systems remains 
sensible
and the 
above mentioned heating effects stay reasonably weak.

In terms of the above introduced quantities, our main analytical
result
is
the prediction 
\begin{equation}
\label{eq:TypTimeEvo}
\< A \>_{\!\rho(t)} 
		= \Auth + |\gamma_t(t)|^2 \left[ \< A \>_{\!\rho_0(t)} - \Auth \right] ,
\end{equation}
where $\Auth = \tr(\rho_{\mic} A)$ is the thermal expectation value introduced below 
Eq.~(\ref{eq:DrivingProtocol:Sin}).
The driving effects are encoded in the response function
$\gamma_\tau(t)$, evaluated at $\tau= t$ in~\eqref{eq:TypTimeEvo},
which is obtained as the solution of the
parametrically $\tau$-dependent family of 
integro-differential equations
\begin{eqnarray}
\dot \gamma_\tau(t)
& = & \int_0^t \!\! \d s\, \gamma_\tau(s) \,\gamma_\tau(t-s) \,
[a_\tau \varVFT(s) + b_\tau  \ddot \varVFT(s)]
\ \ \ 
\label{eq:ResProfEq}
\end{eqnarray}
with initial condition $\gamma_\tau(0) = 1$
and
coefficients
\begin{equation}
\label{eq:ResProfEqCoeffs}
a_\tau := - [F_1(\tau)/\tau]^2\, , \
b_\tau := [F_2(\tau)/\tau-F_1(\tau)/2]^2
\,,
\end{equation}
where
$F_1(\tau):=\int_0^\tau\! \d t \, f(t)$ and $F_2(\tau):=\int_0^\tau\! \d t \, F_1(t)$.
We emphasize that the theory and Eq.~\eqref{eq:ResProfEq} in particular are nonlinear,
which -- in light of the
numerically observed response 
characteristics 
(see Fig.~\ref{fig:Spin5x5}) -- is essential to faithfully reproduce 
the observed behavior.

To derive these results,
we combined and advanced three major theoretical 
methodologies:
(i) a Magnus expansion \cite{bla09} for the propagator $\mc U(t)$
(see below Eq. (\ref{eq:TimeEvo}));
(ii) a mapping of the time-dependent problem~\eqref{eq:H} to 
a parametrically $\tau$-dependent family of 
time-independent auxiliary systems;
(iii) a typicality (or random matrix)
framework \cite{deu91, dab20relax, dab21typical} to 
determine the generic behavior (\ref{eq:TypTimeEvo}) for 
the vast majority of all systems
sharing the same $H_0$, $\varV(E)$, and $f(t)$.
Details of the derivation are collected in the Methods.

Of the adopted
techniques, the Magnus expansion in particular implies
that 
such an approach only
covers the transient dynamics up to a certain maximal time,
which increases as the driving period $T$
becomes smaller.
Since this
maximal time
has been related to the onset of heating
\cite{dal13, dal14, ish18},
the result~\eqref{eq:TypTimeEvo} does not capture 
such heating effects anymore.
Yet it may well remain valid over a quite extended time interval since heating is 
suppressed exponentially for small $T$ 
\cite{mor16, aba17effective, aba15, aba17rigorous, 
pen21, bea21},
see also
Supplementary Note~1
for a more 
detailed discussion of the 
relevant time scales 
and
of the response function $\gamma_\tau(t)$.

Due to the employed typicality framework, in turn, the prediction~\eqref{eq:TypTimeEvo} 
may not reproduce the dynamics accurately in certain setups with strong correlations between 
the observable $A$
and the perturbation $V$.

A more in-depth discussion of the expected regime of applicability is provided in the Methods.

\begin{figure*}
\centering
\includegraphics[scale=1]{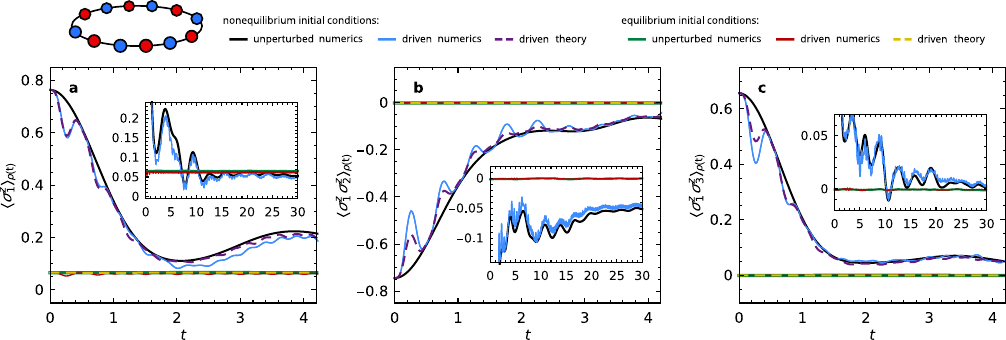} 
\caption{Stalled response in a one-dimensional Ising-type model.
The unperturbed Hamiltonian $H_0$ 
in ~\eqref{eq:H}
is the transverse-field Ising model 
(see sketch), exhibiting
periodic boundary conditions and additional next-nearest-neighbor couplings to break integrability,
$H_0 := -J \sum_{j=1}^L ( \sigma_j^x \sigma_{j+1}^x + \epsilon \, \sigma_j^x \sigma_{j+2}^x + g \, \sigma_j^z)$
with $J = \epsilon = g = \frac{1}{2}$ and $L=24$.
The driving operator is a longitudinal magnetic field,
$V := -J \sum_{j=1}^L \sigma_j^x$.
Time-dependent expectation values $\< A \>_{\!\rho(t)}$ of 
(a) the single-site magnetization $A = \sigma_1^z$,
(b) the nearest-neighbor correlation $A = \sigma_1^z \sigma_2^z$,
and (c) the next-nearest-neighbor correlation $A = \sigma_1^z \sigma_3^z$
are shown for the periodically driven system~\eqref{eq:H}, 
\eqref{eq:DrivingProtocol:Sin}, with driving amplitude $f_0 = 4$ and period $T = 0.5$.
Solid black and blue lines:
numerical results for nonequilibrium initial conditions~\eqref{eq:InitState} with 
$\ket\phi = \ket{\uparrow \downarrow \uparrow \downarrow \cdots }$
(N\'eel state, see sketch), $Q = \id$, $E = -2.4$, and $\Delta E = 1$.
(The corresponding inverse temperature, ground-state energy, and infinite-temperature 
energy are now approximately $0.2$, $-18.5$, and $0$, respectively,
see also above Eq.~\eqref{eq:DrivingProtocol:Sin}.)
Solid green and red lines: same but for equilibrium initial conditions~\eqref{eq:InitState},
i.e., with a Haar-random state $\ket\phi$.
Dashed lines: corresponding theoretical predictions~\eqref{eq:TypTimeEvo},
adopting the numerically obtained unperturbed behavior $\<A\>_{\!\rho_0(t)}$, 
squared response function $\lvert \gamma_t(t) \rvert^2$ (by numerical integration 
of~\eqref{eq:ResProfEq}), and thermal equilibrium
values $\Auth \simeq 0.066,\, 0,\, 0$ in (a), (b), (c), respectively.
Insets: Same numerical data, but with rescaled $x$ and $y$ axes to display the long-time behavior.}
\label{fig:Ising}
\end{figure*}

\begin{figure}
\includegraphics[scale=1]{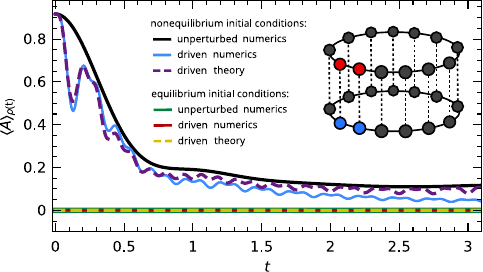}
\caption{Imperfect stalling upon breaking a conservation law.
Time-dependent expectation values $\< A \>_{\!\rho(t)}$ 
of the single-site magnetization $A = \sigma^z_{1,1}$ 
are shown
for a periodically driven 
$2\times 14$ spin double-chain (see inset) 
with Hamiltonian \eqref{eq:H}, \eqref{eq:DrivingProtocol:Sin}, 
\eqref{eq:Spin12x2p:H0}, \eqref{eq:Spin12x2p:V},
and driving period 
$T= 0.25$.
Solid black and blue lines: numerical results for non-equilibrium
initial conditions~(\ref{eq:InitState}) with
$Q = \pi^+_{1,1} \pi^+_{1,2} \pi^-_{2,1} \pi^-_{2,2}$,
for driving amplitudes 
$f_0=0$ (unperturbed, black) and $f_0=3.2$ (driven, blue).
Solid green and red lines: same but for equilibrium
initial conditions~(\ref{eq:InitState}) with $Q = \id$.
Dashed lines: corresponding
theoretical predictions~\eqref{eq:TypTimeEvo}, 
obtained as in Fig.~\ref{fig:Spin5x5} but with
$\Auth = 0$.}
\label{fig:Spin12x2p}
\end{figure}

\subsection*{Interpretation and further examples}

For a quantitative comparison of
our 
theoretical prediction
~\eqref{eq:TypTimeEvo}
to
specific examples, 
some
approximate knowledge of the perturbation 
profile~\eqref{eq:VarV}
is clearly indispensable.
Qualitatively, however, the theory quite remarkably
allows us to make
 some largely general
predictions
without any such specific knowledge.

The first and foremost of these predictions
is based on the general upper bound 
$\lvert \gamma_t(t) \rvert \leq 1$, 
whose detailed analytical derivation is 
provided in
Supplementary Note~7
(see also
Supplementary Note~1.2).
It then immediately follows from \eqref{eq:TypTimeEvo}
that the driving effects are strongly suppressed
whenever \ $\< A \>_{\!\rho_0(t)}\simeq \Auth$,
i.e., whenever the unperturbed system is close to 
thermal equilibrium.
The latter in turn is true for all times $t$ if the
unperturbed system is at thermal equilibrium from the outset,
and for all sufficiently late times $t$ if the 
unperturbed system starts out far from equilibrium 
and is know to thermalize in the long run.
Altogether, our stalled response phenomenon 
is thus analytically predicted to occur under 
very general circumstances.

Next we turn to a more detailed quantitative 
comparison of
the theoretical prediction~\eqref{eq:TypTimeEvo}
with concrete numerical examples.
For the setup considered in
Fig.~\ref{fig:Spin5x5},
exact diagonalization of a smaller system with 
$L = 4$ \cite{dab20modification} 
suggests that the perturbation profile 
$\varV(E)$ from~\eqref{eq:VarV} 
can be approximated very well by an exponential
decay $\varV(0) \, \e^{-\lvert E \rvert / \Delta_v}$.
Utilizing
Ref.~\cite{dab21typical}, 
one moreover finds for the $L=5$ system
in the relevant energy window the numerical estimates 
$\varV(0) \dos \simeq 3.6$ and $\Delta_v \simeq 3.0$,
yielding $\varVFT(t)$ via (\ref{eq:VarVFT}).
All quantities entering the theoretical 
prediction~\eqref{eq:TypTimeEvo}--\eqref{eq:ResProfEq} 
are thus either numerically available [$\< A \>_{\!\rho_0(t)}$, $\Auth$]
or otherwise known [$\varVFT(t)$, $a_\tau$, $b_\tau$], i.e.,
there remains no free fit parameter.

As can be inferred from the solid blue and dashed 
purple lines in Fig.~\ref{fig:Spin5x5},
the theory indeed describes the nontrivial details
of the
driven dynamics
remarkably well.
Notably, it reproduces
the pronounced drop compared to the 
unperturbed curve around 
$t=T/2$ 
and
the quite surprising comeback around 
$t=T$.
Moreover, it indeed also explains the
stalled response behavior 
in Fig.~\ref{fig:Spin5x5} very well,
for initial conditions both close to and far from thermal equilibrium.

Within the framework of Floquet theory,
a related, but distinct effect is well-known under the name ``Floquet prethermalization''
\cite{kuw16, mor16,
aba17effective, aba17rigorous, mal19heating,
mac20, bea21, pen21}:
The dynamics described by the Floquet Hamiltonian approaches a prethermal plateau value
before heating becomes significant and pushes the system towards infinite temperature.
However, the dynamics encoded in the Floquet Hamiltonian only agrees with the actual 
dynamics of the driven system 
stroboscopically, 
i.e.,
only
at integer multiples of the driving period.
A prethermal plateau of the Floquet-Hamiltonian dynamics therefore
still leaves room for strong oscillations of the actual dynamics between the stroboscopic time points where both agree.
Accordingly, the salient new insight provided by our present results is
that no such strong oscillations are observed if the unperturbed system relaxes to 
or starts out from
a thermal equilibrium state.
In other words, our stalled response effect
amounts to a highly nontrivial extension of the established
Floquet prethermalization phenomenon since it means that
the plateau value is 
assumed not only stroboscopically, but even continuously
in $t$.
An extended discussion of the relation between our approach and 
Floquet theory can be found in 
Supplementary Note~4.

As a second example,
we consider a nonintegrable variant of the transverse-field Ising model in 
Fig.~\ref{fig:Ising}, see the figure caption for details.
We particularly emphasize that, for variety and in contrast to Fig.~\ref{fig:Spin5x5},
this setup consists of a one-dimensional system
and globally out-of-equilibrium initial conditions.

Qualitatively, the
numerical results in Fig.~\ref{fig:Ising} once again confirm
the main message of our paper, namely the occurrence of 
stalled response:
Initially, the dynamics shows
a pronounced response when starting away from equilibrium (solid black vs.\ blue lines).
Stalling of that response appears as the unperturbed system approaches thermal equilibrium,
meaning that the oscillations caused by the driving become smaller and smaller.
This is highlighted in the insets, in particular.
(A special feature of this example is that already the unperturbed system
(black lines) 
exhibits a relatively complex and long-lasting relaxation process.)
Likewise, the effects of the driving are barely visible
on the scale of the plot when starting directly 
from a thermal equilibrium state (solid green vs.\ red lines).

For a quantitative comparison of the numerical results 
with
the theoretical prediction~\eqref{eq:TypTimeEvo},
we assume, as in the previous example, 
an approximately exponential perturbation profile $\varV(E) = \varV(0) \e^{-\lvert E \rvert / \Delta_v}$ 
[cf.\ Eq.~\eqref{eq:VarV}],
and use again the theory from Ref.~\cite{dab21typical}
to estimate 
$\varV(0) D_0 \simeq 0.46$ and $\Delta_v \simeq 0.6$.
The resulting 
theoretical
curves in Fig.~\ref{fig:Ising} (dashed lines)
describe the numerics
reasonably well in the initial regime.
In accordance with the discussion below Eq.~\eqref{eq:ResProfEqCoeffs},
for
larger times the theory is no longer quantitatively very accurate (but still
correctly predicts the occurrence of stalling 
per se).
For 
this
reason, no dashed lines are shown in the insets.

Yet another interesting general
prediction of the theory~\eqref{eq:TypTimeEvo} 
(see also beginning of this section)
is that noticeable effects of the driving
(as encoded in $\lvert \gamma_t(t) \rvert^2$)
may actually persist even beyond the relaxation time scale of the unperturbed system
if its long-time expectation value $\Autav := \overline{ \< A \>_{\!\rho_0(t)}}$ 
(infinite time average) differs from the thermal value $\Auth$.
This can happen, for example, if the perturbation $V$ breaks a conservation law of $H_0$.

To verify this prediction, we
consider a third example in Fig.~\ref{fig:Spin12x2p}.
Here the unperturbed system consists of
two isolated spin chains of $L=14$ sites with periodic boundary conditions and Hamiltonian
\begin{equation}
\label{eq:Spin12x2p:H0}
	H_0 := H^{(1)} + H^{(2)}
	\,, \quad
	H^{(i)} := \sum_{j=1}^L \bm\sigma_{i,j} \cdot \bm\sigma_{i,j+1} \, ,
\end{equation}
while the perturbation in (\ref{eq:H})
connects the chains sitewise,
\begin{equation}
\label{eq:Spin12x2p:V}
	V := \sum_{j=1}^L \bm\sigma_{1,j} \cdot \bm\sigma_{2,j} \,;
\end{equation}
see also the sketch in the inset.
The initial state is again of the form~\eqref{eq:InitState} with $E = -14$ and 
$\Delta E = 4$,
restricted to the sector with vanishing
$S^z := \sum_{j} (\sigma^z_{1,j} + \sigma^z_{2,j})$.
(The corresponding inverse temperature, ground state energy, 
and infinite-temperature energy are
now approximately $0.12$, $-50$, and $-1$, respectively
see also above Eq.~\eqref{eq:DrivingProtocol:Sin}.)
However,
for the nonequilibrium setup we now fix two spins in the ``up'' state for 
the first chain and two in the ``down'' state for the second chain
(red and blue, respectively, in the 
sketch), i.e.,
$Q := \pi^+_{1,1} \pi^+_{1,2} \pi^-_{2,1} \pi^-_{2,2}$.
Since the two chains ($i = 1,2$) do not interact in the unperturbed system,
their magnetizations $S_{i}^z := \sum_{j} \sigma^z_{i,j}$
are conserved individually, and thus maintain their initial 
expectation values $2$ and $-2$, respectively, under evolution with $H_0$.
In the driven system, by contrast, only the total $S^z := S^z_1 + S^z_2 $ 
is conserved.
Choosing the single-site magnetization $A = \sigma^z_{1,1}$ as our observable,
we thus find by symmetry that $\Autav = 2/L$ is the long-time 
expectation value of the unperturbed dynamics,
whereas the thermal value of the joint system is $\Auth = 0$.

The numerics in Fig.~\ref{fig:Spin12x2p} (solid blue line)
visualizes the aforementioned imperfect stalling upon breaking a conservation law:
The suppression of the response is the
stronger the closer the unperturbed system is to thermal equilibrium.
Crucially, however, the driving effects still remain visible 
even when the unperturbed dynamics has essentially reached 
its nonthermal long-time value $\Autav$.
Altogether, this confirms the
prediction of~\eqref{eq:TypTimeEvo}
that proximity to thermal equilibrium is indeed  
the decisive condition for stalled response
and not, for example, relaxation of the 
unperturbed system.
Furthermore, this example highlights once again that stalled response 
and Floquet prethermalization are distinct effects:
The present system exhibits Floquet prethermalization,
meaning that the stroboscopic dynamics approaches a stationary plateau,
but no stalled response since $\< A \>_{\!\rho(t)}$ continues to oscillate.

For a quantitative comparison with the theory~\eqref{eq:TypTimeEvo}, we 
again
adopt the same ansatz as before and estimate $\varV(0) \dos = 0.98$ and $\Delta_v = 4.2$
via~\cite{dab21typical}.
The so-obtained prediction~\eqref{eq:TypTimeEvo} (dashed 
purple) agrees rather well with the numerics for 
$t \lesssim 1$.
At later times, the quantitative deviations between the prediction and the numerics increase.
As suggested below~\eqref{eq:ResProfEqCoeffs} and discussed in more detail in the Methods,
we can attribute these deviations to the adopted Magnus expansion and its truncation at second order.
Yet
the above mentioned general qualitative prediction of our theory remains
valid nonetheless.

\subsection*{Basic physical mechanisms}

Intuitively, the basic 
physics behind 
all our above mentioned
numerical and analytical findings 
can
also be understood 
by means of the following simple arguments:
As long as heating is insignificant, we may
focus on
the dynamics within
the initially populated energy interval $\Delta$ (see above~\eqref{eq:VarV}).
Denoting by $P$ the projector onto
the eigenstates $\ketN{\mu}$ with $E_\mu \in \Delta$,
the
Hamiltonian $H(t)$ from (\ref{eq:H})
can thus be reasonably well approximated 
by its
projection/restriction
$\tilde H(t):=PH(t)P$
to
$\Delta$.
Since the microcanonical ensemble $\rho_{\mic}:=P/\tr\{P\}$ commutes with $\tilde H(t)$,
it is a stationary state with respect to $\tilde H(t)$.
Within the present approximation,
a system in thermal equilibrium is thus completely unaffected by the 
periodic driving, and analogously the effects remain weak if the system 
is in a state close to thermal equilibrium.
(Incidentally, the relaxation of a non-equilibrium initial 
state under $\tilde H(t)$ can be heuristically understood by 
similar arguments as in Ref. \cite{laz14equilibrium}.)
On the other hand, subleading effects
like small remnant oscillations and slow heating 
cannot be understood within 
this simplified picture.
Rather, 
these
effects must be attributed to the neglected 
corrections $H(t) - \tilde H(t)$ and,
as a consequence, are
intimately connected with each other.

A complementary, and even 
more simplistic argument is
based on the well-established fact \cite{llo88, gol06, pop06} 
that the vast majority of all pure states with energies 
in $\Delta$ behave akin to $\rho_{\mic}$
for sufficiently large many-body systems.
This 
so-called typicality
property suggests
that once the system has reached 
(or starts out from) such a state,
it remains within this vast majority
in the absence as well as in the presence of the periodic driving.

Essentially, our stalled response 
effect thus seems to be the result of
a 
subtle
interplay
between the system's many-body character and
intriguing peculiarities of thermal equilibrium states.
The above intuitive arguments
moreover suggest 
that the indispensable prerequisites for stalled response
per se
may 
be substantially weaker than those of our
analytical theory
(see also
Supplementary Note~2.4).

\section*{DISCUSSION}

Our core message
is that
the same many-body system may either exhibit a quite 
significant response when 
perturbed by a periodic driving, 
or may not show any notable reaction to the same driving,
depending on whether the unperturbed reference 
system finds itself far from or close to thermal equilibrium.
We demonstrated this
stalled response effect
by numerical examples,
and further substantiated it
by 
sophisticated analytical methods
and by simple physical arguments.

Previous theoretical and experimental studies of periodically driven many-body systems
(e.g., 
Refs.~\cite{dal13, laz14periodic,
laz14equilibrium,
laz15,
pon15manybody,
kuw16, mor16, moe17,
aba17effective,
geo14,
mal19heating, mac20,
pen21, bea21}
among many others)
have been 
very
successful in characterizing the long-term properties of such systems,
including heating effects 
\cite{dal13, dal14, laz14equilibrium, pon15manybody, ish18, mal19heating, ike21, mor21heating}
and their suppression
\cite{aba15, mor16, aba17rigorous, aba17effective, 
pen21, bea21,
geo14,
rus12, laz14periodic, 
laz15,
pon15manybody}.
The latter, in particular, facilitates
the phenomenon of Floquet prethermalization
 \cite{kuw16, mor16,
aba17effective, aba17rigorous, mal19heating, mac20,
pen21, bea21},
a long-lived, stroboscopically quasistationary phase which has been 
exploited, for instance, 
to design various meta 
materials
with promising topological 
and dynamical properties
\cite{
hol16,
moe17, oka19, mac20,
wei21}.

Complementary to those 
long-term features for discrete time points, 
our present focus is on how a many-body system approaches
such prethermal regimes continuously in time.
Overall, we thus arrive at the following 
general picture for periodically driven systems
with moderate driving periods and amplitudes:
Given a thermalizing unperturbed system that 
is prepared sufficiently far from equilibrium,
the periodic perturbations generically lead 
to quite notable response effects 
on short-to-intermediate time scales.
Subsequently, the expectation values approach 
a (nearly) time-independent behavior.
On even much larger time scales, the 
system finally heats up to infinite temperature, 
manifesting itself  in a 
slow drift of the 
expectation values towards their 
genuine
infinite-time limits.

In principle, our 
predictions
can
be readily tested with presently available techniques in, for example, cold-atom \cite{
blo08, blo12, bla12, lan15,
ued20,
geo14} or polarization-echo \cite{geo14,
pen21,
bea21}
experiments.
In practice, previous experimental (as well as theoretical) 
investigations
mostly focused on the long-time behavior and stroboscopic dynamics.
A notable exception 
is the NMR experiment from Ref.~\cite{bea21}:
In Figs.~3(a) and~5(a,b) 
therein,
the NMR signal of the initially out-of-equilibrium system undergoes 
vigorous
oscillations 
at first (called 
``transient approach'' 
in~\cite{bea21}).
Then,
their amplitude gradually decreases as the running mean approaches a 
quasistationary value 
(called ``prethermal plateau'' in \cite{bea21}).
Even later,
the only noticeable effect of the driving is a slow drift as the system heats up 
(called ``unconstrained thermalization'' in \cite{bea21}).
Unfortunately, the available experimental details are not sufficient to compare the measurements 
quantitatively with our 
analytical theory~\eqref{eq:TypTimeEvo}.
Nevertheless,
the observed NMR signal clearly shows the 
general qualitative features of stalled response as predicted by Eq.~\eqref{eq:TypTimeEvo}.

\section*{METHODS}

We first lay out the three main steps in the derivation 
of~\eqref{eq:TypTimeEvo}--\eqref{eq:ResProfEq},
and subsequently address the expected 
validity regime of the employed approximations.

\subsection*{Magnus expansion}

The time evolution of the driven quantum system with Hamiltonian $H(t)$ 
from~\eqref{eq:H} is encoded in the propagator $\mc U(t)$ introduced below Eq.~\eqref{eq:H},
which satisfies the Schr\"odinger-type equation $\frac {\d}{\d t} \mc U(t) = -\I H(t) \mc U(t)$.
Whereas
this equation is formally solved by an (operator-valued) exponential for time-independent Hamiltonians,
no such simple solution is available for the driven case.
To make progress while keeping the setting as general as possible,
we adopt a Magnus expansion \cite{bla09} of the propagator,
writing
\begin{equation}
\label{eq:UMagnus}
	\mc U(t) = \e^{\Omega(t)} \,, 
	\quad
	\Omega(t) = \sum_{k=1}^\infty \Omega_k(t) \, ,
\end{equation}
where the individual terms $\Omega_k(t)$ in the exponent consist 
of integrals over $k-1$ nested commutators of $H(t)$ at different time points.
The virtue of the Magnus series compared to other expansion 
schemes (e.g., a Dyson series) is that $\mc U(t)$ 
remains unitary when truncating~\eqref{eq:UMagnus} at a finite order.

For Hamiltonians of the specific form~\eqref{eq:H},
the first two terms of the general Magnus expansion 
(see, e.g., Ref.~\cite{bla09})
can be readily rewritten as
\begin{subequations}
\label{eq:UMagnus2ndOrder}
\begin{align}
	\Omega_1(t) &= -\I \left[ H_0 t + F_1(t) V \right] , \\
	\Omega_2(t) &= \left[ F_2(t) - \frac{t}{2} F_1(t) \right] [V, H_0] \,,
\end{align}
\end{subequations}
where $[V, H_0] := V H_0 - H_0 V$ 
(commutator), and $F_{1,2}(t)$ are defined 
below Eq.~\eqref{eq:ResProfEqCoeffs}.

\subsection*{Mapping to auxiliary systems}

Adopting the Magnus expansion~\eqref{eq:UMagnus}, the propagator $\mc U(t) = \e^{\Omega(t)}$ 
assumes an exponential form similar to the case of time-independent Hamiltonians.
However, the time dependence of the exponent is generally still complicated.
To proceed, we introduce a one-parameter 
family of time-independent auxiliary Hamiltonians
\begin{equation}
\label{eq:HAux}
	H^{(\tau)} := \I \Omega(\tau) / \tau \,,
\end{equation}
where $\tau > 0$ is treated as an arbitrary but 
fixed parameter.
Starting from the same initial state $\rho(0)$ as in the
actual system of interest, 
any of these Hamiltonians $H^{(\tau)}$ generates a time evolution with the state at time $t$ given by
\begin{equation}
\label{eq:rhoAux}
	\rho(t, \tau) := \e^{-\I H^{(\tau)} t} \rho(0) \e^{\I H^{(\tau)} t} \,.
\end{equation}
Since $\rho(t) = \mc U(t) \rho(0) \mc U(t)^\dagger$, the combination of Eqs.~\eqref{eq:UMagnus}, \eqref{eq:HAux}, and~\eqref{eq:rhoAux} implies that
the state $\rho(t)$ of the driven system of interest coincides with the time-evolved state of the auxiliary system $H^{(t)}$ at time $t$, i.e.,
\begin{equation}
\label{eq:rhoFromAux}
	\rho(t) = \rho(t, t) \,.
\end{equation}
Hence finding the dynamics of the original driven system is equivalent to determining the behavior of all the auxiliary systems with time-independent Hamiltonians $H^{(\tau)}$ up to time $t = \tau$, respectively.

Restricting ourselves to the second order of the Magnus expansion,
we adopt Eqs.~\eqref{eq:UMagnus2ndOrder} in~\eqref{eq:HAux} 
to approximate the auxiliary Hamiltonians as
\begin{equation}
\label{eq:HAux2ndOrder}
	H^{(\tau)}
	\simeq
	H_0 + V^{(\tau)}
\end{equation}
with
\begin{equation}
\label{eq:VAux}
	V^{(\tau)} := \frac{F_1(\tau)}{\tau} V + \left[ \frac{ F_2(\tau)}{\tau} - \frac{F_1(\tau)}{2} \right] \I [V, H_0] \,,
\end{equation}
thereby splitting off the $\tau$-independent reference Hamiltonian $H_0$.

\subsection*{Typicality framework}

It is
empirically well established
that the macroscopically observable behavior of systems with many degrees 
of freedom
can be described by a few effective characteristics despite the vastly complicated 
dynamics of their individual microscopic constituents. 
Detecting and separating the macroscopically relevant properties of a many-body 
system from the intractable microscopic details can arguably be considered as the 
paradigm of statistical mechanics.
The final component of our toolbox to describe the driven many-body dynamics aims at adopting such 
an approach to the  observable expectation values $\< A \>_{\!\rho(t)}$.

To this end, we start with the Hamiltonian $H(t) = H_0 + f(t) V$ from~\eqref{eq:H} 
and temporarily consider an entire class 
(or a so-called ensemble)
of similar driving operators $V$.
Ideally, we would like to establish that all members of such an ensemble exhibit the same observable dynamics.
In practice, what is 
analytically
feasible is a slightly weaker variant of such a statement.
Namely, we demonstrate that nearly all members $V$ of the ensemble
show in very good approximation the same \emph{typical behavior},
and that the fraction of exceptional members, leading to noticeable deviations 
from the typical 
behavior,
is exponentially small in the system's degrees of freedom.

In essence, the defining characteristic of the considered ensembles is the perturbation profile $\varV(E)$ from~\eqref{eq:VarV}.
Introducing the symbol $\av{\,\cdots}$ to denote the average over the $V$ ensemble,
the matrix elements $V_{\mu\nu}$ are treated as independent (apart from the Hermiticity constraint, $V_{\mu\nu} = V_{\nu\mu}^*$) and unbiased ($\av{ V_{\mu\nu}} = 0$) random variables with variance $\av{ \lvert V_{\mu\nu} \rvert^2 } = \varV(E_\mu - E_\nu)$.
Hence the property~\eqref{eq:VarV} of the true 
perturbation is built into the ensemble in an ergodic sense, i.e., upon replacing local averages 
$\lav{\,\cdots}{E}$
(see below Eq.~(\ref{eq:VarV})) 
by ensemble averages $\av{\,\cdots}$.
Due to a generalized central limit theorem (cf.\ Supplementary Note~6),
these first two moments are essentially the only 
relevant characteristics of the $V$ 
ensemble,
i.e., the precise distribution of the $V_{\mu\nu}$ can still take rather general forms.
A detailed definition of the admitted ensembles is provided in Supplementary Note~5.

For time-independent Hamiltonians of the form $H = H_0 + \lambda V$ with a 
constant 
(time-independent) 
perturbation,
it was demonstrated in Refs.~\cite{dab20relax, dab21typical} that those ensembles can
 indeed be employed to predict the observed dynamics in a large variety of settings.
In the following, we will extend the underlying approach to the auxiliary Hamiltonians 
$H^{(\tau)}$ of the form~\eqref{eq:HAux2ndOrder}.
The distribution of the $V_{\mu\nu}$ thus induces a distribution of the matrix elements 
$V_{\mu\nu}^{(\tau)} := \braN{\mu} V^{(\tau)} \ketN{\nu}$ of $V^{(\tau)}$ from~\eqref{eq:VAux}.
In particular, we obtain $\av{ V^{(\tau)}_{\mu\nu} } = 0$ and, together with the 
definitions~\eqref{eq:VarV}, \eqref{eq:ResProfEqCoeffs}, and~\eqref{eq:VAux},
\begin{equation}
\label{eq:VarVAux}
	\av{ \lvert V^{(\tau)}_{\mu\nu} \rvert^2 }
		= -\left[ a_\tau + \left( E_\mu - E_\nu \right)^2 b_\tau \right] \varV(E_\mu - E_\nu) \,.
\end{equation}

As a first step of our typicality argument,
we then calculate the ensemble average $\av{ \< A \>_{\!\rho(t,\tau)} }$ of the time-evolved expectation values.
Deferring the details to 
Supplementary Note~6,
we eventually obtain the relation
\begin{equation}
\label{eq:AvgTimeEvoAux}
	\av{ \< A \>_{\!\rho(t, \tau)} }
		=\Auth + \lvert \gamma_\tau(t) \rvert^2 \left[ \< A \>_{\!\rho_0(t)} - \Auth \right] .
\end{equation}

Here
a Fourier transformation relates
the response function (see above~(\ref{eq:ResProfEq}))
via
\begin{equation}
\label{eq:ResProf}
	\gamma_\tau(t) = \frac{1}{\pi} \lim_{\eta\to 0+} \int \d E \, \e^{\I E t} \, \Im G(E - \I\eta, \tau)
\end{equation}
to the function $G(z, \tau)$,
which solves
\begin{equation}
\label{eq:GAuxEq}
	G(z, \tau) \! \left[ z + \int \! \d E \, D_0 \, G(z \!-\! E, \tau) \! \left( a_\tau - E^2 b_\tau \right) \! \varV(E) \right]
	= 1
\end{equation}
and encodes the ensemble-averaged resolvent of $H^{(\tau)}$ via $\av{ (z - H^{(\tau)})^{-1} } = G(z - H_0, \tau)$.
In
Supplementary Note~7,
we furthermore show that Eqs.~\eqref{eq:ResProf} 
and~\eqref{eq:GAuxEq} imply the relation~\eqref{eq:ResProfEq} for $\gamma_\tau(t)$.

As a next step,
we turn to the deviations $\xi(t, \tau) := \< A \>_{\!\rho(t,\tau)} - \av{ \< A \>_{\!\rho(t, \tau)} }$ between the 
driven dynamics induced by one particular perturbation operator $V$ and the average behavior.
More explicitly, we inspect the probability $\mathbb{P}( \lvert \xi(t, \tau) \rvert \geq x )$ that a randomly 
selected perturbation $V$ generates deviations $\xi(t, \tau)$ that are larger than some threshold $x$.
As explained in more detail in
Supplementary Note~8,
we can find a constant $\delta = 10^{-\mathcal{O}(N_{\mathrm{dof}})}$ (decreasing exponentially with 
the system's degrees of freedom $N_{\mathrm{dof}}$)
such that
\begin{equation}
\label{eq:ProbTimeEvoAuxMain}
	\mathbb{P}( \lvert \xi(t, \tau) \rvert \geq \delta \Delta_A ) \leq \delta \,,
\end{equation}
where $\Delta_A$ is the measurement range of $A$ (difference between its largest and smallest eigenvalues).
In other words,
observing deviations which exceed some exponentially small threshold value
becomes exponentially unlikely as the system size increases,
a phenomenon that is also sometimes 
called  ``concentration of measure'' or ``ergodicity'' 
in the literature.
Consequently,
\begin{equation}
\label{eq:TypTimeEvoAux}
	\< A \>_{\!\rho(t, \tau)} \simeq \av{ \< A \>_{\!\rho(t, \tau)} }
\end{equation}
becomes an excellent approximation for the vast majority of perturbations $V$ in sufficiently large systems.
Combining Eqs.~\eqref{eq:rhoFromAux}, \eqref{eq:AvgTimeEvoAux}, and~\eqref{eq:TypTimeEvoAux},
we thus finally recover our main result~\eqref{eq:TypTimeEvo}.

\subsection*{Limits of applicability}

The class of systems whose Hamiltonian can be written in the form~\eqref{eq:H} is extremely general.
However, the methods described above contain three major
assumptions or idealizations that restrict the types of admissible setups to some extent.

The first issue arises when adopting the Magnus expansion~\eqref{eq:UMagnus} for the propagator $\mc U(t)$.
The question of its convergence is generally a subtle issue and rigorously guaranteed 
in full generality
only up to times 
$t$ such that the operator norm  $\lVert H(s) \rVert$ satisfies $\int_0^t \d s \, \lVert H(s) \rVert < \pi$,
but can extend to considerably longer times in practice nonetheless \cite{bla09}.
Due to the extensive growth of $H(t)$ with the degrees of freedom, 
guaranteed convergence is thus very limited for typical many-body systems,
but the expansion can still remain valuable 
as an asymptotic series for short-to-intermediate times \cite{kuw16, mor16}.
For periodically driven systems in particular,
the (Floquet-)Magnus series amounts to a high-frequency expansion and thus works best for 
small driving periods $T$ \cite{bla09, buk15}.
More generally, the smaller the characteristic time scale of the driving protocol $f(t)$ is,
the larger is the time up to which the expansion offers a satisfactory approximation at any fixed order.

Physically, the breakdown of the Magnus expansion has been related to the onset of heating \cite{dal13, dal14, ish18}.
Generically, many-body systems subject to perpetual driving are expected to
absorb energy indefinitely and heat up to a state of infinite temperature 
\cite{dal14, laz14equilibrium,
pon15manybody, ish18, mal19heating},
unless there are mechanisms preventing thermalization such as an extensive 
number of conserved quantities \cite{rus12, laz14periodic} or many-body 
localization \cite{
laz15,
pon15manybody, aba16}.
Nevertheless, under physically reasonable assumptions about the system, 
such as locality of interactions,
it has been shown that the heating rate is exponentially small in the driving 
frequency \cite{aba15, mor16, aba17rigorous, aba17effective}.
For sufficiently fast driving, therefore,
energy absorption is essentially suppressed for a long time and the Magnus 
expansion can provide a good description of the dynamics.
A more quantitative discussion of the interdependence of the relevant time scales is provided in
Supplementary Note~1.1.

In summary, the Magnus expansion is expected to work as long as the state $\rho(t)$ 
stays roughly within the initially occupied microcanonical energy window $\Delta$ 
of the unperturbed reference Hamiltonian introduced above Eq.~\eqref{eq:VarV}.
Consequently, the stalled-response effect and the applicability of the prediction~\eqref{eq:TypTimeEvo}
are generally expected to persist for longer times at larger initial temperatures
because the relative influence of heating is smaller in this case.
Furthermore, higher temperatures come with a higher density of states,
such that finite-size effects are smaller, too.
The temperature dependence is discussed in more detail in
Supplementary Note~2.2.

A second limitation is our truncation of the Magnus expansion at second order.
In general, this will further restrict applicability towards shorter times and/or faster driving,
but still leaves room for a broad and interesting parameter regime as demonstrated 
examplarily in Figs.~\ref{fig:Spin5x5}--\ref{fig:Spin12x2p}.
In principle, including higher-order terms may be possible, even though it leads to 
severe technical complications in the typicality calculation outlined above
(see also
Supplementary Note~6),
and is thus beyond the scope of our present work.
Besides the response 
function $\gamma_t(t)$,
higher-order corrections are also expected to affect the long-time value ($\Auth$ in Eq.~\eqref{eq:TypTimeEvo}):
It is well known from Floquet theory that this plateau value of Floquet prethermalization
is controlled by the Floquet Hamiltonian \cite{mor16, kuw16, aba17effective, aba17rigorous, mac20}.
The latter agrees with $H_0$ to lowest order,
but can yield different long-time behavior in general,
even though the corrections are generically expected to be small \cite{mor16}.

A third potentially limiting factor for the applicability of our present approach
is the typicality framework,
within which we introduce ensembles of matrix representations $V_{\mu\nu}$ of the 
driving operator
$V$ in the eigenbasis of the reference Hamiltonian $H_0$.
Our main result states that the observable dynamics of nearly all 
members $V$ of
such an ensemble is described by Eqs.~\eqref{eq:TypTimeEvo} 
and \eqref{eq:ResProfEq} (up to the limitiations discussed earlier).
The final point to establish is that the true 
(non-random) driving operator $V$ of actual interest
is one of those typical members of the ensemble,
which evidently requires a faithful modeling of the system's 
most essential
properties with regard to the observable dynamics.

The classes of perturbation ensembles considered here are a compromise 
between what is physically desirable and mathematically feasible.
From a physical point of view,
we would like to emulate the matrix structure of realistic models as closely as possible.
We therefore explicitly incorporate the possibility for sparse (most $V_{\mu\nu}$ are strictly zero) 
and banded (the typical magnitude $\lvert V_{\mu\nu} \rvert$ decays with the energy separation 
$\lvert E_\mu - E_\nu \rvert$ of the coupled levels) perturbation matrices.
These features indeed commonly arise as a consequence of the local and few-body character 
of interactions in realistic systems
as supported by semiclassical arguments \cite{fei89, fyo96}, analytical studies of lattice 
systems \cite{ara16, oli18}, and a large number of numerical examples 
(e.g.\ Refs.~\cite{
beu15, kon15,
jan19}).
Similar assumptions are also well-established 
in random matrix theory and in the context of the 
eigenstate thermalization hypothesis \cite{neu29,deu91,sre94,rig08}.
On the other hand,
the geometry of the underlying model and the structure of interactions 
(for instance their locality) are not explicitly taken into account.
Therefore, the existence of macroscopic transport currents as a consequence of macroscopic 
spatial inhomogeneities
can likely invalidate 
the prediction~\eqref{eq:TypTimeEvo}--\eqref{eq:ResProfEq},
at least for observables $A$ which are sensitive to such
initial spatial imbalances and their equalization
in the course of time.

This is ultimately related to our idealization of statistically
independent matrix elements $V_{\mu\nu}$ for $\mu \leq \nu$.
In any realistic system, some of the matrix elements will certainly mutually depend on each other.
However, it is generally hard to identify (let alone quantify) potential correlations in any given system,
so independence may also be understood as unbiasedness in the absence of more detailed information.
Moreover, mild correlations will often not have a noticeable impact on the properties 
relevant for the observable dynamics \cite{rei19prethermal}.

A specific case where correlations can become relevant, though, are observables $A$ that 
are strongly correlated with the perturbation $V$, most notably if $A = V$.
Since we keep the observable fixed when calculating ensemble averages,
most members of the $V$ ensemble will obviously violate such a special relationship.
Unfortunately, it is not straightforwardly possible to
adapt the method such that the case $A = V$ 
can be described as well because including $A = V$ in the ensemble averages would also affect the 
unperturbed reference dynamics $\< A \>_{\!\rho_0(t)}$.
Numerical explorations and further discussions of this case are provided in 
Supplementary Note~2.4.
Notably, the qualitative predictions of the theory~\eqref{eq:TypTimeEvo} 
and, in particular, the occurrence of stalled response can still be seen 
for the observable $A = V$.

For the rest,
we emphasize that it is
not
necessary for all 
members $V$ of
a certain ensemble to be 
physically realistic.
The decisive question is whether their majority embody the key mechanism underlying the 
observable dynamics in the same way as the true system of interest.
To give an example from textbook statistical mechanics, a large part of states contained
in the canonical ensemble (as a mixed density operator) will be 
unphysical,
and yet its suitability to characterize macroscopically observable properties of closed 
systems in thermal equilibrium is unquestioned provided that the temperature as the 
pertinent macroscopic parameter is chosen appropriately.

More generally, the probabilistic nature of the result implies that any given system can show 
deviations even if all prerequisites are formally fulfilled,
but the probability for such deviations is exponentially suppressed in the system's 
degrees of freedom,
cf.\ Eq.~\eqref{eq:ProbTimeEvoAuxMain}.
For generic many-body systems, we therefore cannot but conclude that 
Eqs.~\eqref{eq:TypTimeEvo}--\eqref{eq:ResProfEq} are expected to hold unless there are specific reasons to the contrary.
The explicit example systems from Figs.~\ref{fig:Spin5x5}--\ref{fig:Spin12x2p} only 
corroborate this observation, noticeably even though the number of degrees of 
freedom is still far from 
being truly macroscopic
in those systems.


\section*{Acknowledgements}

This work was supported in parts by the 
Deutsche Forschungsgemeinschaft (DFG,
German Research Foundation)
within the Research Unit FOR 2692
under Grant No.~355031190 (PR).
We acknowledge support for the publication costs 
by the Open Access Publication Fund of Bielefeld University 
and the Deutsche Forschungsgemeinschaft (DFG).



\clearpage

\onecolumngrid
\begin{center}
\ \\[10pt]
{\large\textbf{SUPPLEMENTARY INFORMATION}}
\end{center}
\vspace{25pt}
\twocolumngrid

\setcounter{subsection}{0}
\renewcommand{\thesubsection}{Supplementary Note~\arabic{subsection}}
\titleformat{\subsection}[display]{\centering\normalfont\bfseries}{\thesubsection:}{3pt}{}
\titleformat{\subsubsection}[block]{\centering\normalfont\em}{\arabic{subsection}.\thesubsubsection.}{5pt}{}

\setcounter{figure}{0}
\renewcommand{\thefigure}{S\arabic{figure}}
\setcounter{equation}{0}
\renewcommand{\theequation}{S\arabic{equation}}
\setcounter{table}{0}
\renewcommand{\thetable}{S\arabic{table}}

Labels of equations and figures in these Supplementary Notes are prefixed by a capital letter ``S'' (e.g., Fig.~S1, Eq.~(S3)).
Any plain labels (e.g., Fig.~1, Eq.~(3), Ref.~[2]) refer to the corresponding items in the main text.

\subsection{Response characteristics}
\label{app:Response}

\subsubsection{Response and heating time scales}
\label{app:Response:TimeScales}

Generically,
a many-body system is expected to absorb energy and thus heat up under periodic driving 
\cite{dal13, dal14, laz14equilibrium, 
pon15manybody, ish18, mal19heating, ike21, mor21heating,hal18}.
As explained in the main paper,
a key prerequisite to observe stalled response
is that these heating effects are sufficiently suppressed
such that the dynamics remains practically confined to the initially occupied (microcanonical) energy shell for the relevant
response and stalling time scales.

\begin{figure}
\includegraphics[scale=1]{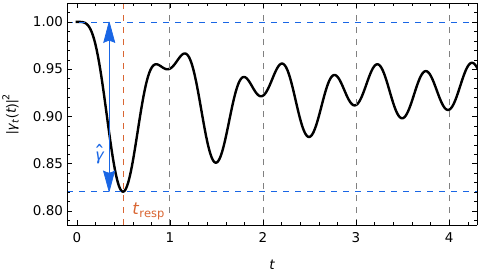}
\caption{Exemplary squared response function $\lvert \gamma_t(t) \rvert^2$ 
(cf.\ \meqref{eq:TypTimeEvo} in the main paper),
obtained by numerical integration of \meqref{eq:ResProfEq} from the main paper for 
an exponential perturbation profile $\tilde v(E) = \e^{-\lvert E \rvert}$ and sinusoidal 
driving $f(t) = \sin(2\pi t)$ (unit period and amplitude).
Dashed gray lines: multiples of the driving period $T = 1$.
Dashed orange line: response time scale $\tResp$, 
defined as the
location of the first minimum of $\lvert \gamma_t(t) \rvert^2$.
Difference between dashed blue lines: response magnitude $\hat\gamma$, cf.\ Eq.~\eqref{eq:ResAmpl}.}
\label{fig:ResProf}
\end{figure}

\begin{figure}
\includegraphics[scale=1]{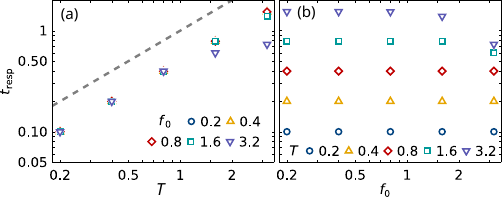}
\caption{Dependence of the response time scale $\tResp$,
defined as the time of the first minimum of $\lvert \gamma_t(t)\rvert^2$,
on the driving period $T$ and amplitude $f_0$
of the sinusoidal driving $f(t) = f_0 \, \sin(2\pi t / T)$
(cf. \meqref{eq:DrivingProtocol:Sin} in the main paper).
(a) $\tResp$ vs.\ $T$ for various $f_0$;
(b) $\tResp$ vs.\ $f_0$ for various $T$.
The function $\lvert \gamma_t(t) \rvert^2$ was evaluated by numerically 
integrating~\meqref{eq:ResProfEq} from the main paper
for an exponential perturbation profile $\tilde v(E) = \e^{-\lvert E \rvert}$. 
Dashed line: $\tResp \propto T$ as a guide to the eye.}
\label{fig:ResTimeScale}
\end{figure}

As a first general guess, the characteristic time scale $\tResp$ of the observable response for a system away 
from equilibrium 
will often be
on the order of the 
characteristic driving time scale $T$
and thus
decrease as $\tResp = \mathrm{O}(T)$ for small $T$.
Within our theory (cf.\ \meqref{eq:TypTimeEvo} in the main paper),
the response is encoded in $\lvert \gamma_t(t) \rvert^2$.
Hence,
$\tResp$ is the typical time scale of the solutions of \meqref{eq:ResProfEq} in the main paper,
an example of which is shown in Fig.~\ref{fig:ResProf}.
For concreteness, we take $\tResp$ to be the time at which the first minimum of $\lvert \gamma_t(t) \rvert^2$ is assumed (see also Fig.~\ref{fig:ResProf})
and illustrate its dependence on $T$
in Fig.~\ref{fig:ResTimeScale}a,
where we indeed observe a linear relationship in the small-$T$ regime.

\begin{figure}
\includegraphics[scale=1]{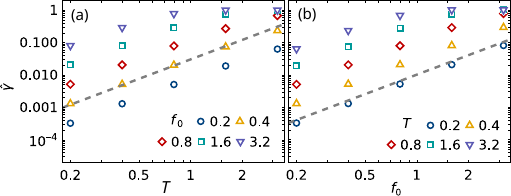}
\caption{Dependence of the response amplitude $\hat\gamma$ 
from Eq.~\eqref{eq:ResAmpl}
on the driving period $T$ and amplitude $f_0$.
(a) $\hat\gamma$ vs.\ $T$ for various $f_0$;
(b) 
$\hat\gamma$ vs.\ $f_0$ for various $T$.
The function $\lvert \gamma_t(t) \rvert^2$ was evaluated by numerically integrating \meqref{eq:ResProfEq} 
from the main paper
for an exponential perturbation profile $\tilde v(E) = \e^{-\lvert E \rvert}$ and sinusoidal driving $f(t) = f_0 \, \sin(2\pi t / T)$.
Dashed lines: $\hat\gamma \propto T^2, f_0^2$ as a guide to the eye.}
\label{fig:ResAmplScale}
\end{figure}

For a system that is initially out of equilibrium,
\emph{stalling} of the observable response
and relaxation to the prethermal plateau are
then predicted to occur on the time scale $\tStall$
on which the dynamics $\< A \>_{\!\rho_0(t)}$ of the associated unperturbed system $H_0$ thermalizes.
Since no assumptions about the unperturbed system are made,
$\tStall$ is largely arbitrary and can vary considerably,
similarly to the relaxation times of isolated many-body systems.
To observe a noticeable effect of the driving and its subsequent stalling,
we 
need $\tResp < \tStall$ (hence $T \lesssim \tStall$).
Moreover,
$\tStall$ must be considerably smaller than the heating time scale $\tHeat$.

Contrary to $\tResp$, this time scale $\tHeat$ for heating
will usually grow as $T$ is decreased.
Specifically,
approximate laws and rigorous upper bounds for the energy absorption rate per degree of freedom, 
$\Gamma \sim \tHeat^{-1}$, have been established in various lattice systems,
for example:
for spins or fermions with local interactions in the linear response regime \cite{aba15} and beyond \cite{aba17effective};
for spins with few-body interactions based on truncated Floquet-Magnus expansions \cite{mor16};
or for hard-core bosons by numerical linked-cluster expansions \cite{mal19heating}.
All those works demonstrate that 
$\Gamma \leq \e^{-\mathrm{O}(1/T)}$
asymptotically for small $T$,
opening up a large initial 
time window
where heating effects are 
insignificant on the typical response and stalling time scales if the driving period is sufficiently small.

Intuitively, one expects that the system can no longer follow the driving for 
$T\to 0$, i.e., the driving effects average out to zero for asymptotically fast driving.
Moreover, the leading order corrections for finite $T$ are expected to be invariant 
under a sign change of $T$, i.e., they generically should scale quadratically with $T$.
Within the theory (cf.\ \meqref{eq:TypTimeEvo} in the main paper),
we can assess the magnitude of the response as the amplitude of $\lvert \gamma_t(t) \rvert^2$ at the first minimum.
Recalling that $\gamma_\tau(0) = 1$,
we thus inspect the quantity
\begin{equation}
\label{eq:ResAmpl}
	\hat\gamma := 1 - \lvert \gamma_{\tResp}(\tResp) \rvert^2
\end{equation}
and find that $\hat\gamma$ indeed
scales quadratically with $T$ for small values,
as shown in Fig.~\ref{fig:ResAmplScale}a.
To achieve a noticeable observable response,
one should thus increase the amplitude $f_0$ of the driving if $T$ is decreased.
As illustrated in Fig.~\ref{fig:ResAmplScale}b,
$\hat\gamma$ likewise scales quadratically with $f_0$ for fixed (sufficiently small) $T$,
whereas the time scale $\tResp$ is largely unaffected by such variations of $f_0$ (cf.\ Fig.~\ref{fig:ResTimeScale}b).
Consequently, a decrease of $T$ can be compensated by a proportional increase of $f_0$ to maintain a similar magnitude of the observable response;
see also
Fig.~\mref{fig:Spin5x5} in the main paper for a visualization in a concrete example system.

The heating rate
$\Gamma \sim \tHeat^{-1}$, 
in turn, also  
grows
quadratically with $f_0$ 
for small amplitudes
within the linear response regime according to Refs.~\cite{mal19heating, ike21, mor21heating}.
However, as mentioned above, the observable response will be very weak if both $f_0$ and $T$ are small.
More precisely,
except for finite-size effects, 
we still expect that the \emph{relative} difference in response between systems near and far from 
thermal equilibrium  will remain significant in this case,
but the stalling effect will be less impressive on an absolute scale.

Hence more interesting is the case of stronger driving beyond the linear response regime.
Here,
general statements
about the heating rate $\Gamma$
are scarce and require more information about the driving operator,
but there is evidence that the dependence of $\Gamma$ on $f_0$ is often nonmonotonic,
such that $\Gamma$ may decrease again eventually as $f_0$ becomes larger \cite{hal18, ike21, mor21heating}.
In any case,
the dependence is typically subexponential.
As $T$ is decreased, and even if $f_0$ is increased accordingly,
the exponential suppression of heating in $1/T$ will thus eventually dominate.
For sufficiently small $T$ and large $f_0$,
we therefore generically expect a regime where heating is insignificant, with a significant 
response away from equilibrium, but strongly suppressed response in 
thermal equilibrium.

In conclusion,
the parameter regime for stalled response essentially coincides with the one for the theoretically and experimentally well-established phenomenon of Floquet prethermalization \cite{kuw16, mor16, els17, aba17effective, mal19heating, mac20, rub20, bea21}
(see also the discussion in the
fifth paragraph of the section 
``Interpretation and further examples'' in the main paper).

\subsubsection{General properties of $\gamma_\tau(t)$}
\label{app:Response:gammaProperties}

Expanding on the discussion of their time scales and amplitudes in the previous subsection,
we collect a few general properties of the solutions $\gamma_\tau(t)$ of the nonlinear integro-differential equation~(\mref{eq:ResProfEq}) in the main paper.
Some of these properties are more easily understood from alternative representations 
of $\gamma_\tau(t)$, such as \meqref{eq:ResProf} in the main paper,
or Eq.~\eqref{eq:gammaDef}, 
defined below in \ref{app:SummaryTypRelaxAvg}.
The equivalence of these representations and \meqref{eq:ResProfEq} in the main paper will be established in \ref{app:ResProfEq} below.
The present section merely serves as a convenient overview.

For all $\tau \in \RR$,
$\gamma_\tau(t)$ satisfies the initial condition
\begin{equation}
	\gamma_\tau(0) = 1
\end{equation}
and is bounded,
\begin{equation}
	\lvert \gamma_\tau(t) \rvert \leq 1 \,,
\end{equation}
see below Eq.~\eqref{eq:ResProfEq:Step3}.
Furthermore, $u(E, \tau)$ from Eq.~\eqref{eq:AvgU2} is real-valued and,
for the considered perturbation ensembles as defined in \ref{app:PerturbationEnsembles},
an even function of $E$.
Since $\gamma_\tau(t)$ is the Fourier transform of $u(E, \tau)$ according to Eq.~\eqref{eq:gammaDef},
it inherits those properties,
i.e.,
\begin{equation}
\label{eq:gamma:RealEven}
	\gamma_\tau(t) \in \RR \,,
	\quad
	\gamma_\tau(-t) = \gamma_\tau(t) \,.
\end{equation}
Similarly, taking for granted that $u(E, \tau)$ is sufficiently regular,
we can conclude that
\begin{equation}
	\lim_{t \to \infty} \gamma_\tau(t) = 0
\end{equation}
for any fixed $\tau$.
We point out, though, that the long-time behavior of $\gamma_\tau(t)$ is of minor interest for our present purposes because,
as discussed in the Methods of the main paper,
the truncated Magnus expansion adopted to derive the theoretical prediction 
from~\meqref{eq:TypTimeEvo} in the main paper
will eventually invalidate it at long times.

\begin{figure*}
\centering
\includegraphics[scale=1]{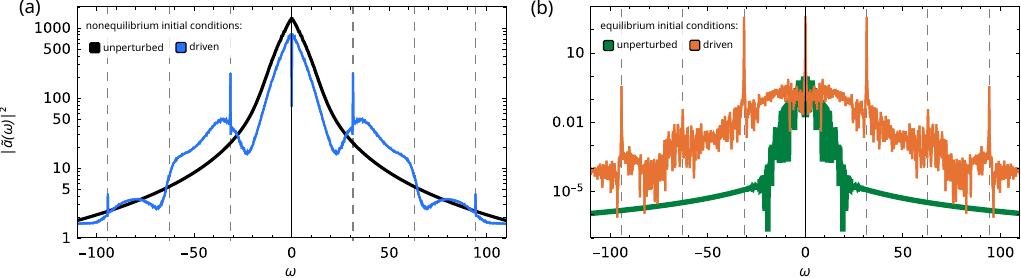} 
\caption{Logarithmically plotted Fourier spectrum $\lvert \tilde a(\omega) \rvert^2$ of the 
nonequilibrium signal $a(t)$ [see Eqs.~\eqref{eq:S:aFT}--\eqref{eq:S:a}]
for the magnetization correlation $A = \sigma^z_{2,2} \sigma^z_{3,3}$ in the two-dimensional spin lattice from Fig.~\mref{fig:Spin5x5} of the main paper
for (a) nonequilibrium and (b) equilibrium initial conditions.
Solid lines: discrete Fourier transformation of the corresponding numerical data in Fig.~\mref{fig:Spin5x5}
as indicated in the legend.
The time series extend up to $t = t_{\max} = 32$ with a step size of $\Delta t = 0.005$,
yielding a frequency resolution $\Delta\omega = \frac{\pi}{16} \approx 0.2$ in the range $\lvert \omega \rvert \leq 200\pi \approx 630$.
Vertical dashed lines: multiples of the driving frequency $\Omega = 2\pi / T = 10 \pi$.}
\label{fig:S:Spin5x5:FT}
\end{figure*}

As for the short-term behavior, we can immediately 
infer from Eq.~\eqref{eq:gamma:RealEven} 
(or from the integro-differential equation~(\mref{eq:ResProfEq}) in the main paper)
that
\begin{equation}
	\dot{\gamma}_\tau(0) = 0 \,.
\end{equation}
More generally, we can expand $\gamma_\tau(t)$ into a Taylor series around $t = 0$.
To this end, it is convenient to introduce the abbreviation $v_\tau(t) := a_\tau v(t) + b_\tau \ddot{v}(t)$,
see also Eq.~\eqref{eq:ResProfEq:VarVFT}.
Denoting by $\gamma_\tau^{(n)}(t)$ the $n$-th derivative with respect to $t$,
a straightforward calulation then yields
the recurrence relation
\begin{equation}
\begin{aligned}
	\gamma_\tau^{(2n)}(0)
		&= \sum_{r=0}^{n-1} \gamma_\tau^{(2n - 2r - 2)}(0) 
		\sum_{k=0}^r \binom{2r}{2k} \gamma_\tau^{(2r - 2k)}(0) v_\tau^{(2k)}(0) 
\end{aligned}
\end{equation}
for the even derivatives,
whereas all odd derivatives vanish
(see also Eq.~\eqref{eq:gamma:RealEven}).
Up to third order in $t$, for example, we thus find
\begin{equation}
\begin{aligned}
	\gamma_\tau(t)
		&= 1 + \frac{v_\tau(0)}{2} t^2 + \mathcal{O}(t^4) 
		\\
		&= 1 - \frac{t^2}{2} \left[ \left( \frac{F_1(\tau)}{\tau} \right)^{\!\!2} v(0) - \left( \frac{F_2(\tau)}{\tau} - \frac{F_1(\tau)}{2} \right)^{\!\!2} \ddot{v}(0) \right] \\
	& \qquad + \mathcal{O}(t^4) \,,
\end{aligned}
\end{equation}
illustrating how the initial decay of $\gamma_\tau(t)$ is controlled by the driving and the decay characteristics of $v(t)$.

\subsection{Additional numerics}
\label{app:Numerics}

\subsubsection{Fourier analysis and nonlinear response}
\label{app:Numerics:Spin5x5:Fourier}

As an interesting complementary, numerical characterization of the system's response 
to the periodic driving,
we consider the Fourier transform
\begin{equation}
\label{eq:S:aFT}
	\tilde a(\omega) := \int_{-\infty}^\infty \d t \, a(t) \, \e^{-\I \omega t}
\end{equation}
of the deviation of the time-evolved expectation values $\<A\>_{\!\rho(t)}$ from the undriven thermal value $\Auth$,
\begin{equation}
\label{eq:S:a}
	a(t) := \< A \>_{\!\rho(t)} - \Auth \,.
\end{equation}
More precisely speaking,
we consider the discrete Fourier transformation of the numerically obtained time series $a(t)$ in an interval $[0, t_{\max}]$ with resolution (step size) $\Delta t$.
For the example from Fig.~\mref{fig:Spin5x5}e in the main paper,
the corresponding Fourier spectra are shown in Fig.~\ref{fig:S:Spin5x5:FT}.

Our first observation is that the periodic driving gives rise to delta peaks 
not only at the driving frequency but also at some of its higher harmonics,
and that the remaining (smooth) part of the perturbed Fourier spectrum 
differs quite notably from its unperturbed counterpart.
(A closer investigation of why the peaks at the second harmonics
seem to be suppressed goes beyond the scope of our present work.)
Both features indicate that the system's response to the periodic driving
is outside the realm of what could be captured by a linear response theory.

Our second observation is that there are no indications of any additional 
delta peaks at noninteger multiples of the driving frequency,
which might have been of interest with respect to the recent
topic of time crystals,
caused by a spontaneous breaking of the discrete time translation 
symmetry \cite{els16, yao17, rus17, zhu19, oje21, kon22, han22}.

\subsubsection{Temperature dependence}
\label{app:Numerics:Spin5x5:Temperature}

\begin{figure*}
\includegraphics[scale=1]{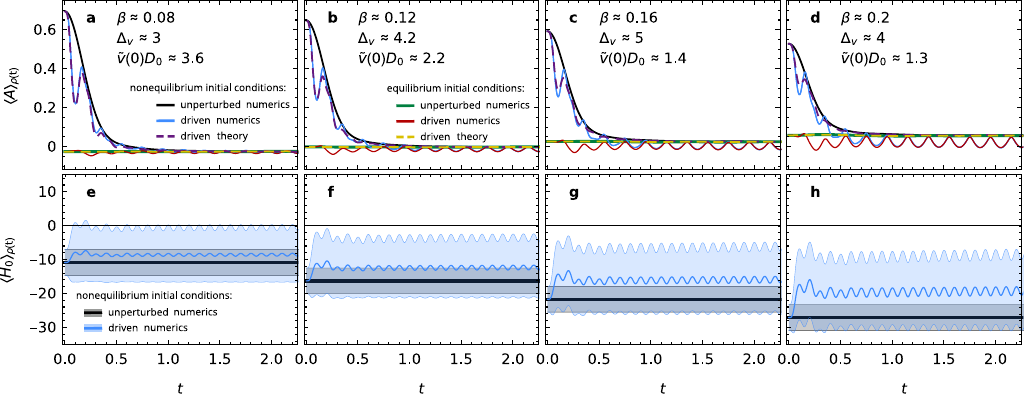}
\caption{Time-dependent expectation values 
for the same
two-dimensional spin lattice 
model as in Fig.~\mref{fig:Spin5x5} of the main paper
with driving period $T = 0.2$ and amplitude $f_0 = 3.2$,
using
initial states of different inverse temperatures $\beta$ as indicated in each column.
Top (a-d): Dynamics of the magnetization correlation $A = \sigma^z_{2,2} \sigma^z_{3,3}$;
bottom (e-h): Corresponding expectation values $\< H_0 \>_{\!\rho(t)}$ of the reference system's energy with bands indicating plus/minus one standard deviation $[ \< H_0^2 \>_{\!\rho(t)} - \< H_0 \>_{\!\rho(t)}^{\,2} ]^{1/2}$.
In (e-h), only results for nonequilibrium initial conditions are shown;
the curves for thermal
equilibrium initial conditions are almost identical.
Parameter values for $\beta \approx 0.08, 0.12, 0.16, 0.2$: $E = -12, -18, -24, -30$ (target energy of the Gaussian filter, see also Fig.~\ref{fig:S:Spin5x5:ObsVsTemp}b);
$\Auth = -0.026, -0.002, 0.026, 0.057$ (see also Fig.~\ref{fig:S:Spin5x5:ObsVsTemp}a);
$\varV(0) D_0 = 3.6, 2.25, 1.44, 1.3$
and
$\Delta_v = 3.0., 4.2, 5.0, 4.0$ 
(see also third paragraph in ``Interpretation and further examples'' of the main paper).}
\label{fig:S:Spin5x5:DiffTemp}
\end{figure*}

\begin{figure}
\includegraphics[scale=1]{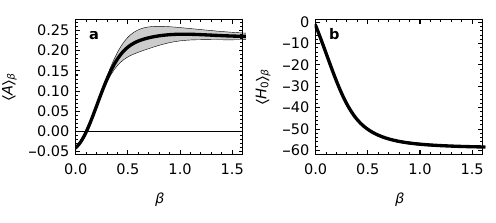}
\caption{Thermal expectation values of $A = \sigma^z_{2,2} \sigma^z_{3,3}$ and $H_0$ in the two-dimensional spin lattice model from Fig.~\mref{fig:Spin5x5} of the main paper 
versus
inverse temperature $\beta$.
Depicted are estimates based on
dynamical typicality methods \cite{bar09, ste14, rei20},
using imaginary-time propagation of three Haar-random states.
Shaded region indicates one standard deviation of the mean
(not visible in b).}
\label{fig:S:Spin5x5:ObsVsTemp}
\end{figure}

In the examples from the main paper (Figs.~\mref{fig:Spin5x5}--\mref{fig:Spin12x2p}),
the energy windows of the initial states were chosen sufficiently far away from the middle of the spectrum
so that heating effects are not trivially absent,
but the corresponding temperatures may still be perceived as relatively high.
The reasons for these choices are mostly of technical nature to mitigate finite-size effects 
(see also \ref{app:Numerics}.3 below).
As briefly mentioned in the main paper (see ``Theory'' and ``Methods'' therein),
the theory is based on the assumption that the initial energy window comprises a large number of energy levels with an approximately homogeneous density of states.
These conditions are commonly satisfied best in the middle of the spectrum 
(highest density of states, smallest relative variations of the density of states),
but the range of compatible temperatures is expected to increase with the system 
size due to the exponential growth of the overall number of levels
as well as of the level density.
Furthermore,
we recall that the stalled-response effect is predicted to occur as long as energy absorption from the periodic driving is negligibile.
The relative influence of this heating is typically smaller for higher temperatures, too,
in the sense that the departure from the initially occupied energy window
is smaller if all other parameters (notably $T$ and $f_0$) are kept fixed (see also Fig.~\ref{fig:S:Spin5x5:DiffTemp}).
Given the limits of our computational resources,
we therefore focused on 
relatively high
temperatures in the main 
paper (see also the subsequent \ref{app:Numerics}.3).

To 
further substantiate
these general arguments,
we discuss the behavior for lower temperatures in Fig.~\ref{fig:S:Spin5x5:DiffTemp}, using the 
same
two-dimensional spin-lattice system 
as in
Fig.~\mref{fig:Spin5x5} of the main 
paper. Likewise,
the initial states are
again
generated according to \meqref{eq:InitState} in the main paper, 
but 
now
using successively lower target energies $E = -12$ 
(similar to Fig.~\mref{fig:Spin5x5}),
$-18, -24, -30$. 
As 
can be inferred from
Fig.~\ref{fig:S:Spin5x5:ObsVsTemp}b,
these correspond to inverse temperatures $\beta \approx 0.08, 0.12, 0.16, 0.2$, respectively.
This Fig.~\ref{fig:S:Spin5x5:ObsVsTemp}b furthermore demonstrates that the 
energy 
in the ground state ($\beta \to\infty$) and at infinite temperature ($\beta =0$) 
are indeed approximately $-60$ and $-1$, respectively,
as stated in the main paper above \meqref{eq:DrivingProtocol:Sin}.

\begin{figure*}[t]
\centering
\includegraphics[scale=1]{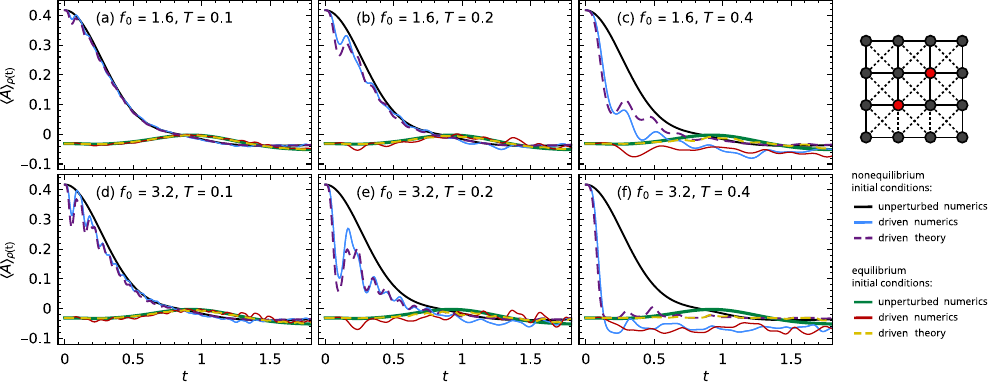} 
\caption{Same as in Fig.~\mref{fig:Spin5x5} of the main paper, 
but now for a $4\times4$ lattice (see also   
\ref{app:Numerics}.3 
for more details).}
\label{fig:S:Spin4x4:Extra}
\end{figure*}

Returning to Fig.~\ref{fig:S:Spin5x5:DiffTemp},
the top panels (a)--(d) show the time-dependent expectation values of the magnetization correlation $A = \sigma^z_{2,2} \sigma^z_{3,3}$ as before.
(Another example -- starting from infinite temperature, $\beta = 0$ -- can be found in 
Fig.~\ref{fig:S:Spin5x5:V2pt:Tinf} below, cf.\ \ref{app:PerturbationProfile2PointFunc}.)
To visualize the energy absorption in the driven system,
we additionally show in the bottom panels (e)--(h) the time-dependent expectation values $\< H_0 \>_{\!\rho(t)}$ of the unperturbed reference Hamiltonian $H_0$ (thick blue line) along with bands of one ``standard deviation'' $[ \< H_0^2 \>_{\!\rho(t)} - \< H_0 \>_{\!\rho(t)}^2 ]^{1/2}$ (blue shaded region).
For comparison, we also indicate the initially occupied energy window (black).

We observe that the driven system is not strictly confined to this 
initially occupied energy window in any of the examples from Fig.~\ref{fig:S:Spin5x5:DiffTemp}.
Instead, all of them show residual heating effects,
which manifest themselves by a drift of the mean energy and broadening of the fluctuations,
as expected for finite driving period.
Notably, however, the departure from the initial energy window is smaller for states of higher temperature (smaller $\beta$).
In line with this observation and the discussion in the main paper,
the observable dynamics $\< A \>_{\!\rho(t)}$ (top panels) shows stronger stalling 
effects for smaller $\beta$, too.
We point out, however, that the response at very early times,
namely the first minimum of $\< A \>_{\!\rho(t)}$ at time $t \approx T/2 = 0.1$,
is strongly suppressed for all displayed temperatures
when comparing the thermal
equilibrium initial conditions to the nonequilibrium behavior.

If all other parameters 
are kept fixed,
stalled response is thus typically more pronounced at higher temperatures,
and most pronounced at early times,
because of relatively weaker heating effects,
similarly as in the case of higher driving frequencies (cf.\ \ref{app:Response}.1).
Stronger suppression at lower temperatures,
in turn, can be expected upon increasing the driving frequency,
or upon increasing the system sizes (see next subsection).

\subsubsection{Finite-size effects}
\label{app:Numerics:FiniteSize}

We recall that our square lattice model from Fig.~\mref{fig:Spin5x5} of the main paper 
only exhibits a relatively 
small extension of $L=5$ sites along each 
of the two spatial directions.
Hence, notable finite-size effects may still be expected.

To get an idea of their relevance,
we consider a smaller version with $L = 4$ of the same
two-dimensional spin-lattice system (cf.\ Eqs.~(\mref{eq:H}), (\mref{eq:Spin5x5:H0}), 
and (\mref{eq:Spin5x5:V}) in the main paper)
as before.
Again, we employ a sinusoidal driving protocol 
(\meqref{eq:DrivingProtocol:Sin} in the main paper) and initial states
as in \meqref{eq:InitState} of the main paper with $Q = \pi^+_{2,2} \pi^+_{3,3}$,
but now choosing $E = -8$ and $\Delta E = 2$
as the target energy window 
in order
to account for the different absolute energy scale of the $L = 4$ system,
and to obtain the same inverse temperature $\beta \approx 0.08$
as in the example with $L=5$.
The thermal expectation value of our observable $A = \sigma^z_{2,2} \sigma^z_{3,3}$ 
in this window now assumes the value $\Auth = -0.040$.

As far as the theoretical prediction 
from Eqs.~(\mref{eq:TypTimeEvo})--(\mref{eq:ResProfEq}) of the main paper 
is concerned,
an advantage of the smaller system size is that we can calculate the perturbation profile 
from~\meqref{eq:VarV} of the main paper directly by exact diagonalization.
We find that it is well approximated by an exponential decay (see also Ref.~\cite{dab20modification}) of the form
\begin{equation}
\label{eq:Spin4x4:VarV}
	\varV(E) = 5.08 \times 10^{-3} \, \e^{-\lvert E \rvert / 8.4}
\end{equation}
for eigenstates of $H_0$ in the relevant energy window, $\lvert E_\mu - E \rvert \leq \Delta E$.
Furthermore, the mean density of states in this window is $D_0 = 425$.
The Fourier transform of $\varV(E)$, cf.\ \meqref{eq:VarVFT} of the main paper, 
is thus
\begin{equation}
\label{eq:Spin4x4:VarVFT}
	\varVFT(t) = \frac{36.3}{1 + (8.4 \, t)^2} \,,
\end{equation}
from which $\gamma_\tau(t)$ can be calculated by integrating \meqref{eq:ResProfEq} 
of the main paper numerically as before.

The so-obtained numerical results along with the corresponding theoretical predictions are 
shown in Fig.~\ref{fig:S:Spin4x4:Extra}.
As a technical aside, we note that in order to avoid additional finite-size artifacts from 
the employed dynamical-typicality method, we averaged over $100$ random 
states $\ket{\phi}$ (cf.\ \meqref{eq:InitState} of the main paper).

\begin{figure*}
\includegraphics[scale=1]{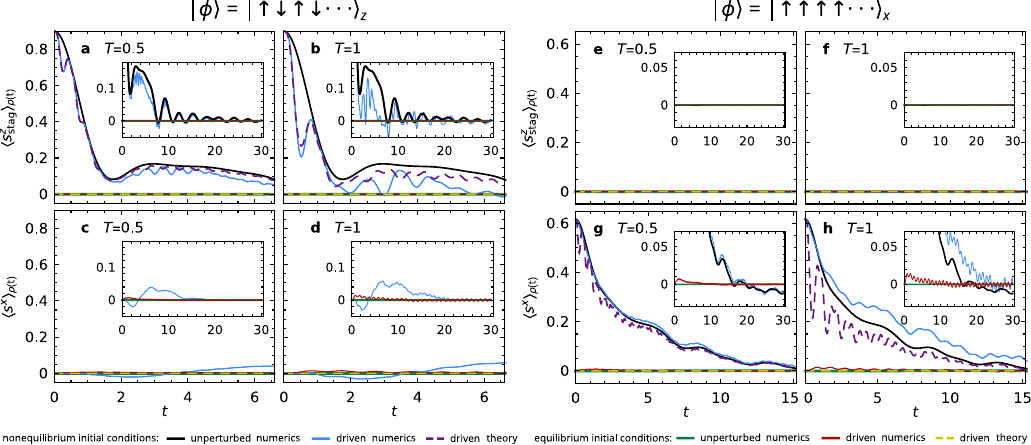}
\caption{Time-dependent expectation values of 
$\Msz$ from Eq.~\eqref{eq:MSz}
(top) and 
$s^x$ from Eq.~\eqref{eq:MSx}
(bottom)
for
the nonintegrable transverse-field Ising model (cf.\ Fig.~\mref{fig:Ising} of the main 
paper)
with
driving amplitude $f_0 = 4$, driving periods $T$
as indicated in each panel, and two different initial states:
N\'eel ordered in the $z$ direction (left, a--d) and fully polarized 
in the $x$ direction (right, e--h).
All other parameters are as in Fig.~\mref{fig:Ising} of the main paper.
Insets: Same numerical data,
but with rescaled $x$ and $y$ axes to display the long-time behavior.
Often, some of the curves are hidden by others,
most notably in panels (e) and (f).}
\label{fig:S:Ising:AeqV}
\end{figure*}

Comparison of the 
these
numerical results for $L = 4$ in Fig.~\ref{fig:S:Spin4x4:Extra} with 
those for $L = 5$ in Fig.~\mref{fig:Spin5x5} of the main paper
provides quite convincing 
evidence
that the stalled response effect should become more pronounced as the system size is 
increased.
In particular,
the small remaining driving 
effects in case of thermal equilibrium initial conditions (red curves)
may be expected to become still 
smaller upon further increasing $L$, which, however, is computationally infeasible for us in practice.
Furthermore,
we
observe that the
theoretical prediction according to \meqref{eq:TypTimeEvo} from the main paper
agrees better with the numerics  (solid lines) for $L = 5$ than for $L=4$,
in accordance with the fact that the derivation of the theory 
assumes large system sizes (see also ``Methods'' in the main paper).

Finally, we also note that increasing $\beta$ (cf.\ \ref{app:Numerics}.2) 
seems to have somewhat similar qualitative 
effects as decreasing the system size $L$. 
For any given such $\beta$,
this suggests once again that the agreement between numerics
and theory, and thus the manifestation of stalled response, 
would significantly improve 
if one increased $L$ further.

\subsubsection{Special case $A = V$}
\label{app:Numerics:AeqV}

As explained in 
the 
``Methods''
of the main paper,
the typicality framework adopted to derive our main analytical result, \meqref{eq:TypTimeEvo} in the main paper,
is not well suited to describe situations in which the observable $A$ is strongly correlated with the driving operator $V$.
To illustrate this and to clarify whether the effect of stalled response per se still applies
(as supported by the heuristic arguments provided in the main paper's ``Basic physical mechanisms'' section),
we investigate the case $A = V$ in the following.
Since $V$ is an extensive operator,
it is appropriate to consider initial states that are globally out of equilibrium
(otherwise 
the observable $A=V$
is expected to exhibit no difference compared to
equilibrium initial states
for asymptotically large systems).
Let us therefore return to our example of the nonintegrable Ising model from Fig.~\mref{fig:Ising} 
of the main paper,
prepared in a (Gaussian-filtered) N\'{e}el state,
\begin{equation}
\label{eq:Ising:InitState}
	\ket\psi \propto \e^{-(H_0 - E)^2 / 4 \Delta E^2} \, 
	\ket\phi
\end{equation}
with $\ket{\phi} = \ket{ \sUp \sDn \sUp \sDn \cdots\, }_z$.
The subscript `$z$' here explicitly indicates that the state is expressed with respect to the spin basis in the $z$ direction.
We also compare the resulting dynamics to the one obtained from the corresponding 
thermal
equilibrium conditions,
emulated as before by choosing $\ket\phi$ to be a Haar-random state in Eq.~\eqref{eq:Ising:InitState}.
In both cases, we furthermore employ in 
Eq.~(\ref{eq:Ising:InitState})
the same parameters $E = -2.4$ and $\Delta E = 1$ as in Fig.~\mref{fig:Ising} of the main paper.

A natural 
observable in view of
the initial state's N\'{e}el order is the staggered 
magnetization 
in the $z$ direction,
\begin{equation}
\label{eq:MSz}
	\Msz := \frac{1}{L}\sum_{j=1}^L (-1)^{j+1} \sigma^z_j \,.
\end{equation}
As anticipated from its close connection to the single-site magnetization $\sigma^z_1$ 
shown in Fig.~\mref{fig:Ising}a of the main paper,
the numerics for $\Msz$ in Fig.~\ref{fig:S:Ising:AeqV}a--b again
exhibits stalled response and good agreement 
with our analytical prediction, \meqref{eq:TypTimeEvo} of the main paper, for sufficiently small times.

Next we turn to Fig.~\ref{fig:S:Ising:AeqV}c--d showing the time evolution of the $x$ magnetization,
\begin{equation}
\label{eq:MSx}
	s^x := \frac{1}{L}\sum_{j=1}^L \sigma^x_j \,.
\end{equation}
Up to a trivial factor $-1/L$, which we henceforth ignore,
this 
observable $A=s^x$ thus coincides with the
driving operator $V$
in our present setup 
(see also the caption of Fig.~\mref{fig:Ising} in the main paper).

Our first observation is that, numerically,
the initial response 
of $s^x$ away from 
equilibrium is  weaker than what is seen in $\Msz$.
(Note that we deliberately chose the same $y$-axis scaling for the plots in panels (a) through (d) to facilitate their direct comparison.)

Second,
the numerically observed
reaction to the driving is significantly stronger away from equilibrium than near 
thermal
equilibrium,
as can be seen by comparing the solid blue and red lines in the insets in particular.
In other words,
the stalled-response phenomenon 
also manifest itself for the special observable $A=V$.

Third, 
the latter applies
despite the fact that the theoretical prediction from \meqref{eq:TypTimeEvo} of the main paper
breaks down for $A=V$, 
as anticipated
there and above.
Indeed, since the thermal expectation value $s^x_{\mathrm{th}} = 0$
and the unperturbed dynamics $\< s^x \>_{\!\rho_0(t)} = 0$ (the associated black 
lines in Fig.~\ref{fig:S:Ising:AeqV}c--d are hidden behind the green ones),
the 
theory from \meqref{eq:TypTimeEvo} of the main paper predicts $\< s^x \>_{\!\rho(t)} = 0$ for the 
driven dynamics, too (the dashed purple lines are likewise hidden behind 
the green solid ones),
while the actually observed numerical response quite notably
deviates from this prediction.
The same theoretical predictions
also apply to the thermal
equilibrium initial conditions,
but here the numerically observed
deviations from 
the theory are less severe due to 
the occurrence of
stalling.

Since the unperturbed dynamics for $A = V$ with N\'{e}el-ordered initial conditions are trivial,
we consider a second example where the initial state is of the same form (cf.\ Eq.~\eqref{eq:Ising:InitState}),
but with $\ket\phi = \ket{\sUp \sUp \cdots\,}_x$ being fully polarized in the $x$ direction.
Hence, we obtain an initial value $\<s^x\>_{\!\rho(0)}$ that is far from the 
thermal equilibrium value $s^x_{\mathrm{th}} = 0$ (see Fig.~\ref{fig:S:Ising:AeqV}g--h).

On the other hand, the unperturbed dynamics of the staggered magnetization is now trivial,
$\< \Msz \>_{\!\rho_0(t)} = (\Msz)_{\mathrm{th}} = 0$,
for both nonequilibrium and equilibrium initial conditions (cf.\ Fig.~\ref{fig:S:Ising:AeqV}e--f; 
black lines hidden behind green ones).
Hence the theory from \meqref{eq:TypTimeEvo} of the main paper predicts $\< \Msz \>_{\!\rho(t)} = 0$ in both cases.
This time, this theoretically predicted behavior is indeed found in the actual dynamics of $\Msz$,
i.e., 
one essentially observes no response
in Fig.~\ref{fig:S:Ising:AeqV}e--f for both the 
nonequilibrium initial state (blue lines hidden behind green ones) and the equilibrium one (red lines).

For 
$A=V$
in Fig.~\ref{fig:S:Ising:AeqV}g--h,
by contrast,
the theory fails to predict the correct dynamics in the nonequilibrium setting (solid blue vs.\ dashed purple lines),
but as emphasized repeatedly,
this is understood to result from the adopted typicality framework in the derivation of \meqref{eq:TypTimeEvo} 
of the main paper.
Notwithstanding,
we observe that the difference between the unperturbed and driven dynamics is overall larger when the system 
is away from 
equilibrium (black and blue lines) compared to the 
situation close to thermal
equilibrium (green and red lines).
Therefore, despite the breakdown of our analytical theory,
the stalled-response effect itself remains observable,
as supported additionally by the heuristic arguments provided 
in the main 
paper's ``Basic physical mechanisms'' section.

\subsection{Relating $v(t)$ to two-point correlation functions}
\label{app:PerturbationProfile2PointFunc}

\begin{figure*}
\includegraphics[scale=1]{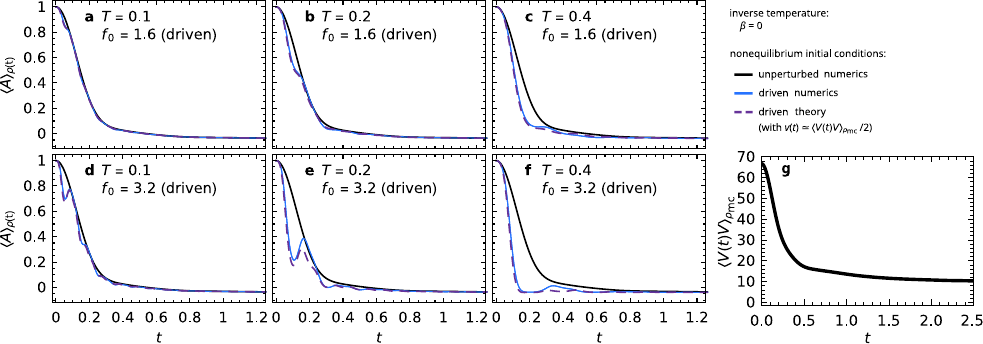}
\caption{(a-f) Time-dependent expectation values $\< A \>_{\!\rho(t)}$ of the magnetization 
correlation $A = \sigma^z_{2,2} \sigma^z_{3,3}$
for
the $5\times 5$ lattice spin system from Eqs.~(\mref{eq:H}), (\mref{eq:Spin5x5:H0}), (\mref{eq:Spin5x5:V}),
(\mref{eq:DrivingProtocol:Sin}) of the main paper (see also Fig.~\mref{fig:Spin5x5} there) at infinite temperature ($\beta = 0$).
Solid lines: numerical results for nonequilibrium initial conditions from Eq.~\eqref{eq:Spin5x5:InitState:Tinf} 
for driving amplitudes $f_0 = 0$ 
(unperturbed, black), 
and for driving periods $T$ and amplitudes $f_0$ as indicated in each panel (driven, blue).
Dashed lines: corresponding theoretical prediction from \meqref{eq:TypTimeEvo} of the main paper,
where $\lvert \gamma_t(t) \rvert^2$ is obtained as the solution of \meqref{eq:ResProfEq} 
of the main paper using the approximation 
from Eq.~\eqref{eq:VarVFT:Approx}
with the numerically obtained two-point 
correlation
function $\< V(t) V \>_{\!\rho_{\mic}}$ from Eq.~\eqref{eq:V2pt:Typ} as shown in (g).}
\label{fig:S:Spin5x5:V2pt:Tinf}
\end{figure*}

\begin{figure*}
\includegraphics[scale=1]{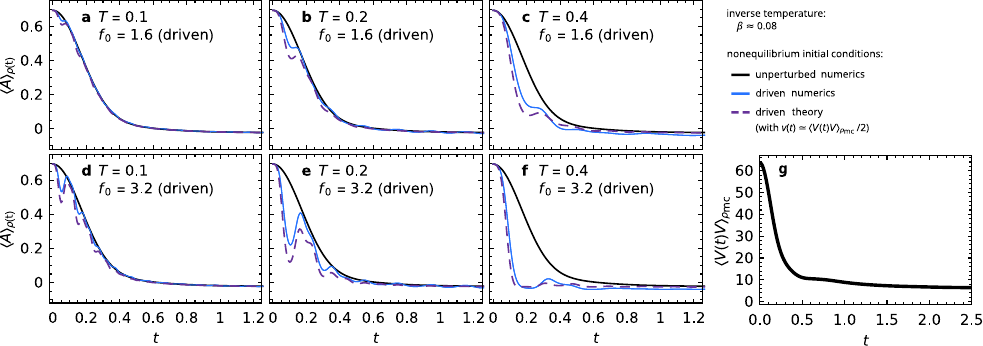}
\caption{Same as in Fig.~\ref{fig:S:Spin5x5:V2pt:Tinf}, but now for 
$\beta \approx 0.08$;
see also
\ref{app:PerturbationProfile2PointFunc} for more details.}
\label{fig:S:Spin5x5:V2pt:E-12}
\end{figure*}

Here, we discuss the connection between the two-point 
correlation
function $\< V(t) V \>_{\!\rho_{\mic}}$ and 
the function
$\varVFT(t)$,
which affects the observable response via \meqref{eq:ResProfEq} of the main paper.
According to its definition in \meqref{eq:VarVFT} of the main paper,
$v(t)$ is the Fourier transform of the perturbation profile $\varV(E) = [\lvert V_{\mu\nu} \rvert^2]_E$ 
from \meqref{eq:VarV} there.
The local energy average $[\lvert V_{\mu\nu} \rvert^2]_E$ as introduced in and around 
this \meqref{eq:VarV} can be expressed more formally as
\begin{equation}
\label{eq:VarV:Explicit}
	[\lvert V_{\mu\nu} \rvert^2]_E
		= \sum_{\mu,\nu \in S_\Delta} k_E(E_\mu, E_\nu) \, \lvert V_{\mu\nu} \rvert^2 \,,
\end{equation}
where $k_E(E_\mu, E_\nu)$ is a suitable averaging kernel
which enforces the condition $\lvert E_\mu - E_\nu \rvert \approx E$ and satisfies $\sum_{\mu, \nu \in S_\Delta} k_E(E_\mu, E_\nu) = 1$.
Moreover, $S_\Delta$ is the set of all indices such that $E_\mu \in \Delta$,
i.e., all states whose energy lie in the initially occupied microcanonical energy window $\Delta$.

Substituting Eq.~\eqref{eq:VarV:Explicit} into \meqref{eq:VarVFT} from the main paper,
we obtain
\begin{equation}
\label{eq:VarVFT:Explicit}
	v(t) = \sum_{\mu, \nu \in S_\Delta} \lvert V_{\mu\nu} \rvert^2 \int \d E \, D_0 \, k_E(E_\mu, E_\nu) \, \e^{\I E t} \,.
\end{equation}
For concreteness, let us now choose $k_E(E_\mu, E_\nu) \simeq \delta(\lvert E_\mu - E_\nu \rvert - E) \,/\, 2 \lvert S_\Delta \rvert D_0$, where $\lvert S_\Delta \rvert = \int_\Delta \d E \, D_0$ is the number of levels in $\Delta$.
Adopting this form in Eq.~\eqref{eq:VarVFT:Explicit}, we obtain
\begin{equation}
	v(t) \simeq \frac{1}{2 \lvert S_\Delta \rvert} \!\sum_{\mu,\nu \in S_\Delta}\! \lvert V_{\mu\nu} \rvert^2 \e^{\I (E_\mu - E_\nu) t}
		= \frac{1}{2 \lvert S_\Delta \rvert} \tr[ P V(t) P V ] \,,
\end{equation}
where $V(t) := \e^{\I H_0 t} V \e^{-\I H_0 t}$ and $P := \sum_{\mu\in S_\Delta} \ket{\mu}\!\bra{\mu}$ is the projector onto $\Delta$.
Observing that the microcanonical ensemble is given by $\rho_{\mic} = P / \lvert S_\Delta \rvert$,
we can finally rewrite $v(t)$ as
\begin{equation}
\label{eq:VarVFT:V2pt}
	v(t) \simeq \frac{1}{2} \< V(t) P V \>_{\!\rho_{\mic}} \,.
\end{equation}
The 
right-hand side is reminiscent of the two-point 
correlation function $\< V(t) V \>_{\!\rho_{\mic}}/2$,
but 
in general not identical
to it because of the additional projector $P$ between the factors of $V$,
which effectively restricts the domain of the matrix product.
However,
the projector $P$ approaches the identity operator as the temperature is increased.

Altogether, these non-rigorous arguments thus suggest 
that it may be possible to employ the approximation
\begin{equation}
v(t)\simeq
\< V(t) V \>_{\!\rho_{\mic}} / 2
\label{eq:VarVFT:Approx}
\end{equation}
at sufficiently high temperatures.

To test this 
conjecture,
we return to the example setup from Fig.~\mref{fig:Spin5x5} of the main paper.
We remove the Gaussian filter in the initial conditions (cf.\ \meqref{eq:InitState} of the main paper),
such that the resulting initial state is effectively at infinite temperature,
\begin{equation}
\label{eq:Spin5x5:InitState:Tinf}
	\ket{\psi} \propto \pi^+_{2,2} \pi^+_{3,3} \ket{\phi} \,,
\end{equation}
where $\ket{\phi}$ is a Haar-random state in the $S^z = -1$ magnetization subsector as before.
The solid lines in Fig.~\ref{fig:S:Spin5x5:V2pt:Tinf}a-f show the corresponding numerically obtained dynamics for the unperturbed (black) and driven (blue) systems.

Furthermore, we numerically calculate the two-point 
correlation function $\< V(t) V \>_{\!\rho_{\mic}}$ at infinite temperature ($\rho_{\mic} = \id / \tr(\id)$)
using dynamical typicality \cite{bar09, ste14, rei20}.
Concretely, we approximate
\begin{equation}
\label{eq:V2pt:Typ}
	\< V(t) V \>_{\!\rho_{\mic}}
		\simeq \< \phi_{\id}(t) | V | \phi_V(t) \> \,,
\end{equation}
where 
$\ket{\phi_Q(t)} := \e^{-\I H_0 t} \ket{\phi_Q}$, $\ket{\phi_Q} := Q \ket{\phi}$, 
and $\ket\phi$ is a Haar-random state as before.
The so-obtained correlation function is shown in Fig.~\ref{fig:S:Spin5x5:V2pt:Tinf}g.

Next we use this result to substitute $v(t) = \< V(t) V \>_{\!\rho_{\mic}} / 2$ in the 
integro-differential equation~(\mref{eq:ResProfEq}) from the main paper for $\gamma_\tau(t)$.
Integrating the equation numerically as before
and utilizing the solution for $\lvert \gamma_t(t) \rvert^2$ in the prediction from \meqref{eq:TypTimeEvo} of the main paper,
we finally obtain the dashed purple lines in Fig.~\ref{fig:S:Spin5x5:V2pt:Tinf}a-f.
The agreement between theory and numerics is remarkably good throughout the inspected range 
of amplitudes and driving periods.

As an
illustration of finite-temperature 
effects in~\eqref{eq:VarVFT:Approx},
we repeat the entire procedure for the identical setup as in Fig.~\mref{fig:Spin5x5} from the main paper,
meaning that the initial state is now again of the same form as in \meqref{eq:InitState} of the main paper with target energy $E = -12$ and width $\Delta E = 4$ (such that $\beta \approx 0.08$).
The corresponding numerical results (solid lines) in Fig.~\ref{fig:S:Spin5x5:V2pt:E-12}a-f are thus identical to those in Fig.~\mref{fig:Spin5x5} of the main paper.
To estimate the time-correlation function $\< V(t) V \>_{\!\rho_{\mic}}$,
we follow the dynamical-typicality approach from Eq.~\eqref{eq:V2pt:Typ} again,
but choose $\ket{\phi_Q} := Q \e^{-(H_0 - E) / 4 \Delta E^2} \ket{\phi}$ now,
thereby emulating the microcanonical density operator of the energy window around the target energy $E = -12$.
This yields Fig.~\ref{fig:S:Spin5x5:V2pt:E-12}g.

The dashed lines in Fig.~\ref{fig:S:Spin5x5:V2pt:E-12}a-f are then again obtained by 
employing
the approximation from Eq.~\eqref{eq:VarVFT:Approx}
in \meqref{eq:ResProfEq} of the main paper.
Comparing theory and numerics,
we observe stronger deviations than in the infinite-temperature case,
but we still recover the qualitative features of the response and even 
achieve reasonable quantitative proximity.

\subsection{Relation to Floquet theory}
\label{app:FloquetTheory}

Numerous insights about the long-time behavior of periodically driven systems,
including heating effects,
metastable plateau regimes (``Floquet prethermalization''),
and their topological properties,
have been obtained using so-called Floquet theory;
see, for example, Refs.~\cite{rus12, dal13, laz14periodic,
dal14, laz14equilibrium,
geo14,
aba15, 
laz15,
pon15manybody,
hol16,
kuw16, mor16, moe17,
aba17effective,
aba17rigorous, 
ish18,
oka19, 
mal19heating, mac20,
pen21, bea21,
wei21,
ike21,
mor21heating,
han22}
and references therein.
As explained in the main paper,
the focus of our present study is on the complementary regime of 
short-to-intermediate times.
Crucially, our methodological approach is distinct from traditional Floquet theory, too.
The following discussion is to clarify their relationship.

Floquet theory is based on the mathematical insight that solutions for the 
propagator $\mc U(t)$ of the time-dependent Schr\"odinger equation 
$\frac{\d}{\d t} \mc U(t) = -\I H(t) \mc U(t)$
can be decomposed as \cite{bla09, tes12, hol16}
\begin{equation}
\label{eq:FloquetAnsatz}
	\mc U(t) = \mc M(t) \, \e^{-\I \HFloq t}
\end{equation}
if $H(t) = H(t + T)$ is time periodic.
Here $\HFloq$ is a time-independent Hermitian operator, the so-called Floquet Hamiltonian.
Furthermore, $\mc M(t)$ is a time-dependent unitary operator of the same periodicity as the driving protocol, $\mc M(t + T) = \mc M(t)$,
sometimes
called the micromotion operator.

Since $\mc U(0) = \id$, one immediately concludes $\mc M(nT) = \id$ for all $n \in \ZZ$
and thus $\mc U(nT) = \e^{-\I \HFloq n T}$.
In other words, the dynamics generated by the time-independent Floquet Hamiltonian $\HFloq$ 
agrees with the dynamics of the actual system (generated by the time-dependent Hamiltonian 
$H(t)$) at integer multiples of the driving period $T$.
(As an aside, the reference time can be chosen arbitrarily,
i.e., by a suitable adaptation of $\HFloq$, one can instead achieve agreement for all $t_n = n T + \theta$ with an arbitrary, but fixed $\theta \in [0, T)$.
Among other things, this implies that the choice of $\HFloq$ is not unique,
but these technical details are not important for the ensuing discussion.)

Exploiting the stroboscopic agreement between the dynamics generated by $\HFloq$ and $H(t)$,
the vast majority of studies investigating the long-time behavior focused on 
properties of the Floquet Hamiltonian $\HFloq$ and largely ignored the periodic 
modulations induced by the micromotion operator $\mc M(t)$.
For example, ``Floquet prethermalization'' \cite{kuw16, mor16,
aba17effective, aba17rigorous, mal19heating,
mac20, bea21, pen21} describes the observation that,
for sufficiently small driving periods $T$,
the dynamics generated by $\HFloq$ resembles ordinary ``prethermalization'' \cite{lan16, mor18}:
The system relaxes from its initial nonequilibrium state to some quasistationary intermediate state,
and the true thermal equilibrium state is only approached at much later times.
In the Floquet case,
the quasistationary intermediate state arises as a result of the strong suppression of heating at fast driving.
Crucially, the system can thus spend long times close to this nontrivial intermediate state,
before heating eventually takes it towards the featureless infinite-temperature (``thermal equilibrium'') state.

The periodic modulations by the micromotion operator $\mathcal{M}(t)$ are usually disregarded when discussing this effect.
Nevertheless, $\mathcal{M}(t)$ can generally still induce a strong time dependence of observable expectation values,
even if the stroboscopic dynamics generated by $\HFloq$ relaxes to a plateau value.
An example is provided in Fig.~\mref{fig:Spin12x2p} of the main paper,
where the stroboscopic dynamics becomes stationary,
but $\< A \>_{\!\rho(t)}$ continues to oscillate.

Our principal observation,
the phenomenon of stalled response,
implies
that those periodic modulations are also suppressed if the accompanying unperturbed system 
finds itself near thermal equilibrium and heating is negligible.
Put differently,
the micromotion operator $\mathcal{M}(t)$ affects states far away from 
equilibrium more strongly than states close to thermal equilibrium.
Coming back to the example from Fig.~\mref{fig:Spin12x2p} of the main paper once again,
we recall that the unperturbed system there relaxes to an equilibrium state.
Crucially, however, this state is different from the thermal equilibrium state of the full system,
and therefore $\mathcal{M}(t)$ continues to have an effect even though the $\HFloq$ 
dynamics has settled down.

From a technical point of view,
the reason why we can characterize the response continuously in time rather than stroboscopically
is that we do \emph{not} adopt a decomposition like in Eq.~\eqref{eq:FloquetAnsatz}.
Instead, we always work with the full propagator $
\mc U(t)$,
particularly when employing the Magnus expansion according to \meqref{eq:UMagnus} in the main paper.

On the other hand,
the fact that we truncate the Magnus expansion at second order
means that our characterization of the ``prethermal'' plateau state is less accurate than state-of-the-art results obtained from high-frequency expansions of stroboscopic Hamiltonians (such as the Floquet Hamiltonian) \cite{mor16, kuw16, aba17effective, aba17rigorous, mac20}.
Within our approximation,
and if we tacitly assume that $f(t)$ averages to zero over one driving period,
the plateau is essentially determined by the time-averaged Hamiltonian $H_0$,
see also the ``Limits of applicability'' subsection in the Methods.
This is primarily relevant for the long-time expectation value $\Auth$ in the prediction 
from \meqref{eq:TypTimeEvo} of the main paper.
In light of the literature on stroboscopic dynamics and observing that $H_0$ is the first-order approximation of $\HFloq$,
higher-order corrections could shift this value.
On the other hand, it was argued in Ref.~\cite{mor16} that these corrections will generically be small and that the prethermal plateau state is well approximated by the microcanonical ensemble of $H_0$.
In any case, assuming such a shift in \meqref{eq:TypTimeEvo} from the main paper
would give room for small remnant oscillations between the plateau value prescribed by the Floquet Hamiltonian and the thermal value approached by the unperturbed dynamics $\< A \>_{\!\rho_0(t)}$.
We emphasize, however, that higher-order corrections will also change the ``response function'' $\gamma_\tau(t)$ and may even affect the overall structure of the prediction.

\subsection{Driving-operator ensembles}
\label{app:PerturbationEnsembles}

The typicality approach employed in this work
(cf.\ Methods in the main paper)
covers 
statistical ensembles of driving operators $V$ of the following general form:
The distributions are expressed in terms of 
probability densities for
the matrix representation $V_{\mu\nu} := \braN{\mu} V \ketN{\nu}$ in the eigenbasis of the 
reference Hamiltonian $H_0$.
These matrix elements are assumed to be statistically
independent apart from Hermiticity ($V_{\mu\nu} = V_{\nu\mu}^*$).
The probability density
$p(V)$ of $V$ with respect to the Lebesgue measure 
$[\d V] := [ \prod_\mu \d V_{\mu\mu} ] [ \prod_{\mu < \nu} 2 \d(\Re V_{\mu\nu}) \d(\Im V_{\mu\nu}) ]$
can therefore be written in the form
\begin{equation}
\label{eq:VPDF}
	p(V) = \prod\nolimits_{\mu \leq \nu} p_{\mu\nu}(V_{\mu\nu})
\end{equation}
with $p_{\mu\nu}(v) := \av{ \delta(v - V_{\mu\nu}) }$.
(We recall that $\av{\,\cdots}$ denotes the average over the $V$ ensemble.)
The marginal probability densities are of the form $p_{\mu\nu}(v) = p_{\lvert E_\mu - E_\nu \rvert}(v)$,
where $\{ p_E(v) \}_{E>0}$ is a family of probability 
densities on $\CC$ with mean zero and variance $\varV(E)$ (cf.\ \meqref{eq:VarV} in the main paper),
and $p_0(v)$ is a probability density on $\RR$ 
with mean zero
and variance $\varV(0)$.
We furthermore assume that the statistical properties of the $V_{\mu\nu}$ are 
unbiased with respect to the choice of the (unphysical) phase of the eigenvectors $\ketN{\mu}$,
meaning that $p_E(v)$ only depends on the absolute value $\lvert v \rvert$.

Note that this 
automatically implies the vanishing mean for $E > 0$ ($\mu \neq \nu$).
The assumption of a vanishing mean of the distribution $p_0(v)$ (i.e., of the diagonal matrix elements), 
in turn, constitutes no loss of generality because any bias could be gauged away by adding a 
constant (proportional to the identity) to $H_0$, which does not alter the dynamics.
In light of the generalized central limit theorem effective below (see \ref{app:SummaryTypRelaxAvg} 
and the discussion below Eq.~\eqref{eq:lambda} in particular),
the considered classes include essentially all reasonable, unbiased 
distributions $p_E(v)$ for the matrix elements $V_{\mu\nu}$
which are compatible with the perturbation profile 
$\varV(E)$ from \meqref{eq:VarV} of the main paper.

In terms of the distribution $p(V)$ from Eq.~\eqref{eq:VPDF}, 
the ensemble average $\av{\,\cdots}$ can thus be written explicitly as
\begin{equation}
\label{eq:av}
	\av{\,\cdots} \equiv \int [\d V] \,\cdots\, p(V) \,.
\end{equation}

\subsection{Ensemble-averaged auxiliary dynamics}
\label{app:SummaryTypRelaxAvg}

In 
the following,
we evaluate the ensemble-averaged auxiliary dynamics $\av{ \< A \>_{\!\rho(t, \tau)} }$,
i.e., we establish Eqs.~(\mref{eq:AvgTimeEvoAux})--(\mref{eq:GAuxEq}) from the main paper.
We focus on an arbitrary but fixed $\tau>0$.
The
time-dependent expectation values of the observable $A$ in the state $\rho(t, \tau)$, 
obtained by evolving $\rho(0)$ for the time $t$ with the auxiliary Hamiltonian 
$H^{(\tau)}$ from Eqs.~(\mref{eq:HAux2ndOrder}) and (\mref{eq:VAux}) of the main paper, can then be written as
\begin{equation}
\label{eq:ATimeEvoAux}
\begin{aligned}
	\< A \>_{\!\rho(t, \tau)}
		&= \sum_{\substack{ \mu_1, \mu_2, \\ \nu_1, \nu_2}} \rho_{\mu_1 \nu_2}(0) A_{\mu_2 \nu_1} \\
			& \quad \times \sum_{m, n} \e^{\I (E^{(\tau)}_n - E^{(\tau)}_m) t} 
			U^{(\tau)}_{m \mu_1} U^{(\tau)}_{n \mu_2}  U^{(\tau)*}_{m \nu_1}  U^{(\tau)*}_{n \nu_2} \,.
\end{aligned}
\end{equation}
Here $E^{(\tau)}_n$ denotes the eigenvalue of $H^{(\tau)}$ corresponding to the eigenvector $\ket{n(\tau)}$
and $U^{(\tau)}_{n \mu} := \langle n(\tau)\ketN{\mu}$
are the overlaps between those
$\ket{n(\tau)}$
and the eigenvectors $\ketN{\mu}$ of the unperturbed reference Hamiltonian $H_0$.
Hence,  the evaluation of $\av{ \< A \>_{\!\rho(t, \tau) } }$ requires calculating ensemble 
averages over four factors of eigenvector overlaps $U_{n\mu}^{(\tau)}$.
In addition, averages over eight such factors will be needed later in 
\ref{app:SummaryTypRelaxVar} to determine the average 
$\av{ \xi(t,\tau)^2 }$ of the fluctuations $\xi(t, \tau)$, cf.\ above \meqref{eq:ProbTimeEvoAuxMain} in the main paper.
By a suitable extension of the methods developed in the Supplemental Material of Ref.~\cite{dab20relax},
these fourth and eighth moments can be traced back to combinations of the second moment
\begin{equation}
\label{eq:AvgU2}
	\av{ \lvert U^{(\tau)}_{n\mu} \rvert^2 } =: u(E_n - E_\mu, \tau)
\end{equation}
with the function $u(E, \tau)$ still to be determined.
Adopting those results,
one finds that the ensemble-averaged auxiliary dynamics satisfies
\begin{equation}
\label{eq:S:AvgTimeEvoAux}
	\av{ \< A \>_{\!\rho(t, \tau)} }
		= \< A \>_{\!\tilde\rho(\tau)} + 
		\lvert \gamma_\tau(t) \rvert^2 
		\left[ \< A \>_{\!\rho_0(t)} - \< A \>_{\!\tilde\rho(\tau)} \right] ,
\end{equation}
where
\begin{align}
\label{eq:gammaDef}
	\gamma_\tau(t) &:= \int \d E \, D_0 \, \e^{\I E t} \, u(E, \tau) \,, \\
\label{eq:rhoTildeDef}
	\tilde\rho(\tau) &:= \sum_{\mu, \nu} \rho_{\mu\mu}(0) \, \tilde u(E_\mu - E_\nu, \tau) \ketN{\nu}\braN{\nu} \,, \\
\label{eq:uTildeDef}
	\tilde u(E, \tau) &:= \int \d E' \, D_0 \, u(E - E', \tau) \, u(E', \tau) \,.
\end{align}
Here $D_0$ is the density of states of the reference Hamiltonian 
$H_0$ in the energy window $\Delta$ as introduced above \meqref{eq:VarV} of the main paper.

The remaining step thus consists of evaluating and interpreting the second moment 
$\av{ \lvert U^{(\tau)}_{n\mu} \rvert^2 } = u(E_n - E_\mu, \tau)$ from Eq.~\eqref{eq:AvgU2}.
Introducing the resolvent or Green's function
\begin{equation}
\label{eq:GDef}
	\mathcal G^{(\tau)}(z) := (z - H^{(\tau)})^{-1}
\end{equation}
of $H^{(\tau)}$,
this second moment can be written as \cite{mir00, haa10}
\begin{equation}
\label{eq:U2FromG}
	\av{ \lvert U^{(\tau)}_{n\mu} \rvert^2 }
		= \frac{\lim\limits_{\eta\to 0+} \! \avv{ \mc G_{\mu\mu}^{(\tau)}(E^{(\tau)}_n \!-\! \I\eta) - \mc G_{\mu\mu}^{(\tau)}(E^{(\tau)}_n \!+\! \I\eta) }}{2\pi\I D_0}
\end{equation}
with $\mc G_{\mu\nu}^{(\tau)}(z) := \braN{\mu} \mc G^{(\tau)}(z) \ketN{\nu}$.
We focus on a Hilbert space of finite dimension $N \gg 1$ (e.g., the energy window from above \meqref{eq:VarV} 
in the main paper).
Employing so-called supersymmetry methods \cite{ver85, efe96, mir00, haa10, ber10},
those matrix elements $\mc G_{\mu\nu}^{(\tau)}(z)$
of the resolvent can be expressed as a Gaussian integral over commuting (complex-valued) variables $x_\alpha$ and anticommuting (Grassmann) variables $\chi_\alpha$ ($\alpha = 1, \ldots, N$),
which we collect in a supervector $X := (X_1^\T \; \cdots \; X_{N}^\T)^\T$ with $X_\alpha := (x_\alpha \; \chi_\alpha)^\T$.
Denoting the associated integration measure by $[\d X \d X^*] := \prod_\alpha \d x_\alpha \d x_\alpha^* \d\chi_\alpha \d\chi_\alpha^*$ and defining the diagonal matrices $L^\pm := \operatorname{diag}(\pm 1, 1)$ as well as the shorthand $z^\pm := E \pm \I \eta$ with $E \in \RR$, $\eta > 0$,
we can then write
\begin{equation}
\label{eq:GGaussianSuperIntegral}
	\mc G^{(\tau)}_{\mu\nu}(z^\pm) = \mp \I \int \frac{ [\d X \d X^*] }{ (\mp 2\pi)^{N} }  x_\mu x_\nu^* \, \e^{\I X^\dagger \left[ (z^\pm - H^{(\tau)}) \otimes L^\pm \right] X } .
\end{equation}
To average over the perturbation ensemble,
we inspect the part of the exponent in Supplementary 
Eq.~\eqref{eq:GGaussianSuperIntegral} that depends on the random variables $V_{\mu\nu}$.
Employing the definition from \meqref{eq:VAux} of the main paper,
we obtain
\begin{widetext}
\begin{equation}
\label{eq:GRandExponent}
	\e^{-\I X^\dagger (V^{(\tau)} \otimes L^\pm) X}
		= \e^{-\I \sum_\alpha \lambda_{\alpha\alpha}^{(\tau)}  X_\alpha^\dagger L^\pm X_\alpha V_{\alpha\alpha}
			-\I \sum_{\alpha < \beta} (\Re V_{\alpha\beta}) (\lambda_{\alpha\beta}^{(\tau)} X_\alpha^\dagger L^\pm X_\beta + \mathrm{c.c.})
			- \I \sum_{\alpha < \beta} \I (\Im V_{\alpha\beta}) (\lambda_{\alpha\beta}^{(\tau)} X_\alpha^\dagger L^\pm X_\beta - \mathrm{c.c.} ) }
\end{equation}
\end{widetext}
where ``$\mathrm{c.c.}$'' indicates the complex conjugate of the preceding term and
\begin{equation}
\label{eq:lambda}
	\lambda_{\alpha\beta}^{(\tau)}
		:= \frac{F_1(\tau)}{\tau} - \I (E_\alpha - E_\beta) \left[ \frac{F_2(\tau)}{\tau} - \frac{F_1(\tau)}{2} \right] .
\end{equation}
Recalling the definition of the perturbation ensembles from \ref{app:PerturbationEnsembles},
the exponent in Eq.~\eqref{eq:GRandExponent} 
is thus a sum of $N^2$ independent and unbiased random variables.
Adopting the central limit theorem,
this sum approaches an unbiased normal distribution as $N \to \infty$,
whose variance is given by the sum of variances of the individual terms,
regardless of further details of the distributions $p_E(v)$ from below Eq.~\eqref{eq:VPDF}.
For sufficiently large $N$,
we can therefore take all $p_E(v)$ to be, for example, Gaussian distributions with mean zero and variance $\varV(E)$
because they lead to the same limiting distribution in Eq.~\eqref{eq:GRandExponent},
i.e., we can adopt, without loss of generality,
\begin{equation}
	p_0(v) = \frac{\e^{-v^2 / 2 \varV(0)}}{\sqrt{2\pi \varV(0)}}
	\quad \text{and} \quad
	p_{E>0}(v) = \frac{ \e^{-\lvert v \rvert^2 / \varV(E)} }{ \pi \varV(E) }
	\,.
\end{equation}
Evaluating the ensemble average of Eq.~\eqref{eq:GRandExponent} according to 
Eq.~\eqref{eq:av} then amounts to computing a product of $N^2$ one-dimensional Gaussian integrals.
Substituting the result into Eq.~\eqref{eq:GGaussianSuperIntegral} yields
\begin{widetext}
\begin{equation}
\label{eq:GAvg1}
	\av{ \mc G^{(\tau)}_{\mu\nu}(z^\pm) }
		= \mp \I \int \frac{ [\d X \d X^*] }{ (\mp 2\pi)^{N} }  x_\mu x_\nu^* \, \e^{-\frac{1}{2} \sum_{\alpha, \beta} \lvert \lambda_{\alpha\beta}^{(\tau)} \rvert^2 \varV(E_\alpha \!-\! E_\beta) \str(X_\alpha X_\alpha^\dagger L^\pm X_\beta X_\beta^\dagger L^\pm) + \I \sum_\alpha (z^\pm - E_\alpha) X_\alpha^\dagger L^\pm X_\alpha } ,
\end{equation}
where $\str(\,\cdots)$ denotes the supertrace.
Next we perform a supersymmetric Hubbard-Stratonovich transformation \cite{efe96, haa10} to rewrite the exponential of the fourth-order term in $X$ as a superintegral involving only quadratic terms in $X$,
namely
\begin{equation}
\label{eq:HSTrafo}
	\e^{-\frac{1}{2} \sum_{\alpha, \beta} s_{\alpha\beta} \str( X_\alpha X_\alpha^\dagger L^\pm X_\beta X_\beta^\dagger L^\pm) }
		= \int \frac{[\d R]}{(2\pi)^N} \e^{ -\frac{1}{2} \sum_{\alpha, \beta} (s^{-1})_{\alpha\beta} \str(R_\alpha R_\beta) + \I \sum_\alpha \str(R_\alpha X_\alpha X_\alpha^\dagger L^\pm) } \,,	
\end{equation}
where $s$ denotes the symmetric matrix with $s_{\alpha\beta} = \lvert \lambda_{\alpha\beta}^{(\tau)} \rvert^2 \varV(E_\alpha \!-\! E_\beta)$ and $s^{-1}$ is its inverse.
Furthermore,
the auxiliary $(2\times 2)$ supermatrices $R_\alpha$ are parametrized as
\begin{equation}
	R_\alpha := \lmat r_{1\alpha} & \rho_\alpha \\ \rho_\alpha^* & \I r_{2\alpha} \rmat
\end{equation}
with real-valued $r_{1\alpha}$, $r_{2\alpha}$ and Grassmann numbers $\rho_\alpha$, $\rho_\alpha^*$,
and $[\d R] := \prod_\alpha \d R_\alpha$ with $\d R_\alpha := \d r_{1\alpha} \d r_{2\alpha} \d\rho_\alpha \d\rho_\alpha^*$ for short.
Substituting Eq.~\eqref{eq:HSTrafo} into Eq.~\eqref{eq:GAvg1} allows us to carry out the resulting Gaussian superintegral over $X$,
leading to
\begin{equation}
\label{eq:GAvg2}
	\av{ \mc G^{(\tau)}_{\mu\nu}(z^\pm) }
		= \delta_{\mu\nu} \int \frac{[\d R]}{(2\pi)^N} \, \left[ (R_\mu + z^\pm - E_\mu)^{-1} \right]_{11} \, \e^{-\str\left[ \frac{1}{2} \sum_{\alpha, \beta} (s^{-1})_{\alpha\beta} R_\alpha R_\beta + \sum_\alpha \ln( R_\alpha + z^\pm - E_\alpha) \right] } \,.
\end{equation}
\end{widetext}
To calculate the remaining integral over the supermatrices $R_\alpha$,
we adopt a saddle-point approximation,
exploiting that the exponent of the integrand in Eq.~\eqref{eq:GAvg2} is extensive in $N \gg 1$ and thus dominated by the region around the highest saddle points of the exponent in the complex, multi-dimensional $R$ plane along a suitably chosen integration contour.
To find this stationary point,
we look for supermatrices $R_\mu$ such that the first variation of the exponent in 
Eq.~\eqref{eq:GAvg2} with respect to $R$ vanishes, i.e.,
\begin{equation}
\label{eq:SaddlePointEqMat}
	R_\mu + \sum_\alpha s_{\mu\alpha} \, (R_\alpha + z^\pm - E_\alpha)^{-1} = 0 \,.
\end{equation}
From the possibly multiple solutions of this saddle-point equation,
we have to select the dominant one that can be reached by a deformation of the original integration contour without crossing any singularities.
The saddle-point approximation of Eq.~\eqref{eq:GAvg2} is then obtained as the product of the integrand and the inverse square root of the superdeterminant corresponding to the second variation 
of the exponent in Eq.~\eqref{eq:GAvg2},
where both are evaluated at the dominating saddle point.
Since the integrand is invariant under (pseudo)unitary transformations $R_\mu \mapsto T R_\mu T^{-1}$ with fixed $T$ satisfying $T^\dagger L^\pm T = L^\pm$,
further solutions can be generated from any given one $\hat R_\mu$ via $T \hat R_\mu T^{-1}$.
Focusing on diagonal solutions first,
the matrix-valued Eq.~\eqref{eq:SaddlePointEqMat} decouples into two identical equations for its entries.
Consequently, any diagonal solution will be of the form $\hat R_\mu = \hat r(E_\mu, z^\pm, \tau) \id$ for some scalar function $\hat r(E_\mu, z^\pm, \tau)$ (explicitly indicating the dependence on $E_\mu$, $z^\pm$, and $\tau$ again)
such that
\begin{equation}
\label{eq:SaddlePointEqScalar}
	\hat r(E_\mu, z^\pm, \tau) + \sum_\alpha \frac{ s_{\mu\alpha} }{ z^\pm - E_\alpha + \hat r(E_\alpha, z^\pm, \tau) } = 0 \,.
\end{equation}
Since $\hat R_\mu$ is proportional to the identity matrix,
$T \hat R_\mu T^{-1} = \hat R_\mu$ and the equivalent solutions collapse back onto the diagonal one.
Moreover, since the superdeterminant of any matrix proportional to $\id$ is unity,
the contribution involving the second variation of the exponent in Eq.~\eqref{eq:GAvg2} amounts to a trivial factor of one.
After the saddle-point approximation, we thus find
\begin{equation}
\label{eq:GAvg3}
	\av{ \mc G^{(\tau)}_{\mu\nu}(z^\pm) } = \frac{\delta_{\mu\nu}}{ z^\pm - E_\mu + \hat r(E_\mu, z^\pm, \tau) } \,.
\end{equation}
Finally, we rewrite the sum in Eq.~\eqref{eq:SaddlePointEqScalar} as an integral over the density of states $D_0$ and exploit that the latter is approximately constant within the relevant energy window (see above \meqref{eq:VarV} in the main paper).
We also substitute $s_{\alpha\beta} = \lvert \lambda^{(\tau)}_{\alpha\beta} \rvert^2 \varV(E_\alpha - E_\beta)$ 
as defined below Eq.~\eqref{eq:HSTrafo}.
Adopting Eqs.~\eqref{eq:lambda} and~\meqref{eq:ResProfEqCoeffs},
we can express it more explicitly as $s_{\alpha\beta} = [-a_\tau + (E_\alpha - E_\beta)^2 b_\tau] \varV(E_\alpha - E_\beta)$.
This quantity is homogeneous in energy in that it only depends on the difference $E_\alpha - E_\beta$.
Due to the homogeneous density of states,
$\hat r(E_\mu, z^\pm, \tau)$ will thus only depend on the difference $z^\pm - E_\mu$ (and $\tau$), too,
i.e., $\hat r(E_\mu, z^\pm, \tau) = \hat r(z^\pm - E_\mu, \tau)$.
Consequently, Eq.~\eqref{eq:SaddlePointEqScalar} takes the form
\begin{equation}
\label{eq:SaddlePointEqScalar2}
	\hat r(z^\pm - E, \tau) - \int \d E' \, D_0 \frac{ a_\tau - (E - E')^2 b_\tau }{ z^\pm - E' + \hat r(z^\pm - E', \tau) } = 0
\end{equation}
Finally, we define the function
\begin{equation}
\label{eq:GAuxDef}
	G(z^\pm, \tau) := [ z^\pm + \hat r(z^\pm, \tau) ]^{-1} \,.
\end{equation}
Substituting into Eq.~\eqref{eq:SaddlePointEqScalar} and shifting the integration variable,
we conclude that
$G(z, \tau)$ satisfies \meqref{eq:GAuxEq} from the main paper.
Moreover, we can exploit Eqs.~\eqref{eq:GDef}, \eqref{eq:GAvg3}, and~\eqref{eq:GAuxDef} to confirm that
\begin{equation}
\label{eq:S:RelationGAvgG}
\begin{aligned}
	\av{ \mc G^{(\tau)}_{\mu\nu}(z^\pm) } 
		&= \av{\braN{\mu} (z^\pm - H^{(\tau)})^{-1} \ketN{\nu}} \\
		&= \delta_{\mu\nu} G(z^\pm - E_\mu, \tau)
\end{aligned}
\end{equation}
as stated below \meqref{eq:GAuxEq} in the main paper.
Furthermore,
Eqs.~\eqref{eq:AvgU2} and~\eqref{eq:U2FromG} imply that
$u(E, \tau) = \lim_{\eta\to 0+} \Im G(E - \I\eta, \tau) / \pi D_0 $.
Together with Eq.~\eqref{eq:gammaDef}, this establishes \meqref{eq:ResProf} from the main paper.

Lastly, we exploit that the typical scale of $u(E, \tau)$ as a function of $E$ is much larger than
the level spacing.
This can be roughly understood by observing that $u(E, \tau)$ quantifies how strongly 
eigenvectors of $H^{(\tau)}$ and $H_0$ are mixed by the perturbation $V^{(\tau)}$,
cf.\ Eq.~\eqref{eq:AvgU2}.
Hence the energy scale of $u(E, \tau)$ is reciprocal to the response time scale of the system.
Therefore, a natural response time necessitates that the energy scale is much larger 
than the exponentially small level spacing in macroscopic systems.
A more detailed justification will be provided in \ref{app:SummaryTypRelaxVar} 
(starting below Eq.~\eqref{eq:VarTimeEvoAux2}).
Crucially, this entails the same property for $\tilde u(E, \tau)$ from Eq.~\eqref{eq:uTildeDef}
and thus implies that $\tilde\rho(\tau)$ from Eq.~\eqref{eq:rhoTildeDef} 
resembles a microcanonical density operator \cite{deu91, rei21}.
Even if the initial state occupies only a few energy levels of the reference Hamiltonian,
the perturbation $V^{(\tau)}$ typically spreads these populations out across
a large number of neighboring states.
Moreover, if the reference system satisfies the eigenstate thermalization hypothesis (ETH) \cite{deu91, sre96, sre99, rig08}
then the expectation value $\< A \>_{\!\bar\rho}$ in the diagonal ensemble $\bar\rho = \sum_\mu \rho_{\mu\mu}(0) \ketN{\mu}\braN{\mu}$ already coincides with the thermal equilibrium value $\Auth$ (see below \meqref{eq:DrivingProtocol:Sin} in the main paper),
and this effect will only be reinforced by the additional averaging caused by the convolution with $\tilde u(E, \tau)$ in 
Eq.~\eqref{eq:rhoTildeDef}.
Hence we conclude that $\< A \>_{\!\tilde\rho(\tau)} = \Auth$ for all practical purposes in 
Eq.~\eqref{eq:S:AvgTimeEvoAux},
which establishes \meqref{eq:AvgTimeEvoAux} from the main paper.

\subsection{Integro-differential representation of $\gamma_\tau(t)$}
\label{app:ResProfEq}

In this
section, we derive the relation from \meqref{eq:ResProfEq} of the main paper for the response 
function $\gamma_\tau(t)$,
which was defined in
Eq.~\eqref{eq:gammaDef}
and was subsequently shown in \ref{app:SummaryTypRelaxAvg} 
[see remarks below Eq.~\eqref{eq:S:RelationGAvgG}]
to be equivalent to \meqref{eq:ResProf} of the main paper.

Our starting point is
this representation of $\gamma_\tau(t)$ from \meqref{eq:ResProf}  of the main paper,
which involves the ensemble-averaged resolvent $G(z, \tau)$ of $H^{(\tau)}$
defined in Eq.~\eqref{eq:GAuxDef}.
Below this equation,
it was also established that the latter function $G(z, \tau)$
is the solution of the nonlinear integral equation~(\mref{eq:GAuxEq}) from the main paper.
Writing $z = x - \I\eta$ with fixed $\eta > 0$,
we multiply both sides of~\meqref{eq:GAuxEq} in the main paper by $\e^{\I x t}$ and integrate over $x \in \RR$.
Defining
\begin{equation}
\label{eq:ResProfEq:hAux}
	h_\eta(t, \tau) := \int \d x \; \e^{\I x t} \, G(x - \I\eta, \tau) \,,
\end{equation}
this yields
\begin{widetext}
\begin{equation}
\label{eq:ResProfEq:Step1}
	-\I \left( \frac{\partial}{\partial t} + \eta \right) h_\eta(t, \tau) + \int \d E \, D_0 \, \varV(E, \tau) \int \d x \int \d y \; \e^{-\I x t} \, G(x - \I\eta, \tau) \, G(y - \I\eta, \tau) \, \delta(x - y - E) = 2 \pi \delta(t) \,,
\end{equation}
\end{widetext}
where
\begin{equation}
\label{eq:ResProfEq:VarV}
	\varV(E, \tau) := \left[ a_\tau - E^2 b_\tau  \right] \varV(E) \,.
\end{equation}
By analogy with~\meqref{eq:VarVFT} in the main paper,
we also introduce the corresponding Fourier transform
\begin{equation}
\label{eq:ResProfEq:VarVFT}
	\varVFT(t, \tau) := \int {\d E}\,D_0 \, \varV(E,\tau) \, \e^{\I E t}  \,.
\end{equation}
Next, we employ the Fourier identity $2 \pi \delta(x) = \int \d s \, \e^{\I x s}$ to replace $\delta(x - y - E)$ on the left-hand side of Eq.~\eqref{eq:ResProfEq:Step1} by an integral.
Exploiting the definitions from Eqs.~\eqref{eq:ResProfEq:VarVFT} and~\eqref{eq:ResProfEq:hAux},
we take the limit $\eta \to 0+$ and find
\begin{equation}
\label{eq:ResProfEq:Step2}
\begin{aligned}
	& \frac{\partial h_{0+}(t, \tau)}{\partial t}
		- \int \frac{\d s}{2\pi\I} \, h_{0+}(t-s, \tau) \, h_{0+}(s, \tau) \, \varVFT(s, \tau) \\
		& \; = 2 \pi \I \, \delta(t) \,.
\end{aligned}
\end{equation}
In view of the $\delta$ inhomogeneity on the right-hand side,
we now make an ansatz of the form $h_{0+}(t, \tau) = 2\pi\I \, \Theta(t) \, \tilde\gamma_\tau(t)$,
where $\Theta(t)$ is the Heaviside step function and $\tilde\gamma_\tau(t)$ is assumed to be a bounded and sufficiently smooth function of $t$.
Substituting into Eq.~\eqref{eq:ResProfEq:Step2},
we obtain
\begin{equation}
	\Theta(t) \, \frac{\partial \tilde\gamma_\tau(t)}{\partial t} - \int_0^t \! \d s \; \tilde\gamma_\tau(t-s) \, \tilde\gamma_\tau(s) \, \varVFT(s, \tau) = 0 \,.
\end{equation}
Combining the relation~(\mref{eq:ResProf}) from the main paper and Eq.~\eqref{eq:ResProfEq:hAux},
we find that $\gamma_\tau(t) = [h_{0+}(t,\tau) - h_{0+}(-t, \tau)^*]/2\pi\I$
and thus
\begin{equation}
 \gamma_\tau(t) = \Theta(t) \, \tilde\gamma_\tau(t) + \Theta(-t) [\tilde\gamma_\tau(-t)]^* \,.
\end{equation}
For $t > 0$, it follows that $\gamma_\tau(t)$ satisfies the integro-differential equation
\begin{equation}
\label{eq:ResProfEq:Step3}
	\frac{\partial\gamma_\tau(t)}{\partial t} = \int_0^t \! \d s \; \gamma_\tau(t-s) \, \gamma_\tau(s) \, \varVFT(s, \tau) \,.
\end{equation}
Likewise, for $t < 0$, Eq.~\eqref{eq:ResProfEq:Step3} follows by observing that $\varV(E, \tau)$ 
from Eq.~\eqref{eq:ResProfEq:VarV} is real-valued and even in $E$,
implying the same for the Fourier transform $\varVFT(s, \tau)$ from 
Eq.~\eqref{eq:ResProfEq:VarVFT} as a function of $s$.
Furthermore, the initial condition $\gamma_\tau(0) = 1$ as stated below~\meqref{eq:ResProfEq} in the main paper
and the bound $\lvert \gamma_\tau(t) \rvert \leq 1$ are understood from 
Eq.~\eqref{eq:gammaDef} and the 
definition of $u(E, \tau)$ from Eq.~\eqref{eq:AvgU2}
by observing that $\int \d E \, D_0 \, u(E, \tau) \simeq \av{ \sum_n \lvert U_{n\mu}^{(\tau)} \rvert^2 } = 1$.

Finally, upon substitution of Eq.~\eqref{eq:ResProfEq:VarV} 
into~\eqref{eq:ResProfEq:VarVFT} and partial integration,
we find that
\begin{equation}
	\varVFT(s, \tau) = \left[ a_\tau +  b_\tau \frac{\d^2}{\d s^2} \right] \varVFT(s) \,,
\end{equation}
confirming
that Eq.~\eqref{eq:ResProfEq:Step3} is equivalent to \meqref{eq:ResProfEq} from the main paper.

\subsection{Ensemble variance and fluctuations of auxiliary dynamics}
\label{app:SummaryTypRelaxVar}

In this 
section,
we establish the bound~(\mref{eq:ProbTimeEvoAuxMain}) from the main paper,
which provides the concentration property that promotes the ensemble 
average from \meqref{eq:AvgTimeEvoAux} of the main paper (see also \ref{app:SummaryTypRelaxAvg}) 
to a prediction for nearly all individual perturbation operators in the ensemble, i.e., \meqref{eq:TypTimeEvoAux} 
of the main paper.

We first analyze the variance $\av{ \xi(t, \tau)^2 }$ of the deviations 
$\xi(t, \tau) = \< A \>_{\!\rho(t, \tau)} - \av{ \< A \>_{\!\rho(t, \tau)} }$ between 
the dynamics for a particular perturbation and the ensemble-averaged behavior.
From Eq.~\eqref{eq:ATimeEvoAux},
we understand that $\xi(t, \tau)^2$ involves eight factors of the basis 
transformation matrix elements $U_{n\mu}^{(\tau)} = \< n(\tau) \mu \>_{\!0}$.
Similarly as in \ref{app:SummaryTypRelaxAvg},
the pertinent ensemble average can be broken down into contributions involving only the 
second-moment characteristic $u(E, \tau)$ 
from Eq.~\eqref{eq:AvgU2}
by adopting the methods of Ref.~\cite{dab20relax}.
This yields
the upper bound
\begin{equation}
\label{eq:VarTimeEvoAux2}
	\av{ \xi(t, \tau)^2 } \leq c \, \Delta_A^{\,2} \max_E u(E, \tau) \,,
\end{equation}
where $\Delta_A$ is the measurement range of $A$ (largest minus smallest eigenvalue) 
and $c$ is a positive constant of order $10^3$ or less, independent of any system 
details (particularly $H_0$, $f(t)$, $\varV(E)$, $\rho(0)$, or $A$).

To understand how this bound scales with the system size,
we take a closer look at the function $u(E, \tau)$.
According to its definition in Eq.~\eqref{eq:AvgU2},
$u(E, \tau)$ quantifies how much an eigenvector $\ketN{\mu}$ of the reference Hamiltonian $H_0$ contributes to the eigenvector $\ket{n(\tau)}$ of $H^{(\tau)}$ that lies a distance $E = E_n - E_\mu$ away from it in the spectrum.
Since the level density in generic many-body systems scales exponentially with the degrees of freedom $\dof$,
any 
driving operator
$V$ with a noticeable effect on the auxiliary Hamiltonian $H^{(\tau)}$ from \meqref{eq:HAux2ndOrder} 
of the main paper will inevitably mix a large number $N_v$ of energy levels \cite{dab20relax}, i.e.,
\begin{equation}
\label{eq:NumMixedLevels}
	N_v = 10^{\mc O(\dof)} \,.
\end{equation}
Taking for granted that $u(E, \tau)$ is reasonably smooth,
this implies that it will typically extend across a scale of order $N_v / D_0$ in $E$ and that its maximum will be at most of order $N_v^{-1}$,
i.e., $\max_E u(E, \tau) \lesssim N_v^{-1}$.
Introducing a suitable constant
\begin{equation}
\label{eq:delta}
	\delta \sim (c N_v^{-1})^{1/3} = 10^{-\mathcal{O}(\dof)} \,,
\end{equation}
we can thus rewrite Eq.~\eqref{eq:VarTimeEvoAux2} as
\begin{equation}
\label{eq:VarTimeEvoAux1}
	\av{ \xi(t, \tau)^2 } \leq \delta^3 \, \Delta_A^{\,2} \,.
\end{equation}

To reinforce Eq.~\eqref{eq:NumMixedLevels} further, we also note that
perturbations mixing only a small number of energy levels will only induce changes of the dynamics on time scales of order $D_0$ because they only affect the corresponding frequencies $E_n^{(\tau)} - E_m^{(\tau)} \sim D_0^{-1}$ in 
Eq.~\eqref{eq:ATimeEvoAux}.
Considering the extremely large level density, the time scale associated with $D_0$ is unimaginably large and typically exceeds the age of the universe by many orders of magnitude.
For all times of practical interest,
the dynamics under $H^{(\tau)}$ would thus be indistinguishable from the reference dynamics under $H_0$.
Observing that $\gamma_\tau(t) \approx 1$ for these times according to Eq.~\eqref{eq:gammaDef} and the normalization of $u(E, \tau)$,
this limiting case is reflected correctly in Eq.~\eqref{eq:S:AvgTimeEvoAux},
meaning that perturbations violating Eq.~\eqref{eq:NumMixedLevels} 
are covered by the final result as well, despite being physically uninteresting.

Exploiting Chebyshev's inequality,
we can bound the probability that the deviations $\lvert \xi(t, \tau) \rvert$
exceed a predefined threshold $x$ in terms of the variance,
\begin{equation}
\label{eq:ProbTimeEvoAux}
	\mathbb{P}\!\left( \lvert \xi(t, \tau) \rvert \geq x \right) \leq \frac{\av{ \xi(t, \tau)^2 }}{x^2} \,.
\end{equation}
Utilizing Eq.~\eqref{eq:VarTimeEvoAux1} and choosing $x = \delta \Delta_A$,
we then obtain the relation~(\mref{eq:ProbTimeEvoAuxMain}) from the main paper.

\end{document}